\documentclass[nofootinbib,aps,prd,superscriptaddress,preprintnumbers,%
 showpacs,showkeys,floatfix,amssymb,amsfonts]{revtex4-1}

\usepackage{arydshln}
\usepackage{graphicx}
\usepackage{bm}
\usepackage{amsmath}
\usepackage{amssymb}
\usepackage{amsfonts}
\usepackage{float}
\usepackage{dsfont}  
\usepackage{slashed}  
\usepackage{booktabs}
\usepackage{multirow}
\usepackage{subfigure}
\usepackage[sort&compress]{natbib}
\usepackage{ulem}
\usepackage{empheq}
\usepackage{color}
\usepackage{makecell}

\newcommand{\be}{\begin{equation}}  
\newcommand{\ee}{\end{equation}}  
\newcommand{\beq}{\begin{eqnarray}} 
\newcommand{\eeq}{\end{eqnarray}}

\newcommand{\bea}{\begin{eqnarray}}
\newcommand{\eea}{\end{eqnarray}}

\newcommand{\MSb}{{\overline{\rm MS}}}

\usepackage{xr-hyper}

\usepackage[svgnames,x11names,table]{xcolor}
\usepackage[colorlinks=true,linkcolor=MediumBlue,citecolor=MediumBlue,urlcolor=MediumBlue]{hyperref}
\usepackage{array}
\usepackage{hyperref}

\begin{document}

\title{Mellin Moments of Pion and Kaon Unpolarized PDFs \\[1ex] from Nonlocal Operators in Lattice QCD}
\author{Joshua Miller}
\email{joshua.miller0007@temple.edu}
\affiliation{Department of Physics,  Temple University,  Philadelphia,  PA 19122 - 1801,  USA}

\author{Joseph Torsiello}
\affiliation{Department of Physics,  Temple University,  Philadelphia,  PA 19122 - 1801,  USA}

\author{Krzysztof Cichy}
\affiliation{Faculty of Physics and Astronomy, Adam Mickiewicz University, ul.\ Uniwersytetu Pozna\'nskiego 2, 61-614 Pozna\'{n}, Poland}

\author{Martha Constantinou}
\email{marthac@temple.edu}
\affiliation{Department of Physics,  Temple University,  Philadelphia,  PA 19122 - 1801,  USA} 

\author{Joseph Delmar}
\affiliation{Department of Physics,  Temple University,  Philadelphia,  PA 19122 - 1801,  USA}
\affiliation{Physics Division, Argonne National Laboratory, Lemont, IL 60439, USA\\[5ex]}

\begin{abstract}

We present a first-principles lattice-QCD determination of Mellin moments of the unpolarized pion and kaon parton distribution functions using matrix elements of boosted mesons coupled to nonlocal operators containing a straight Wilson line. The calculation is performed on an $N_f=2+1+1$ ensemble of maximally twisted-mass fermions with a clover term, with lattice volume $32^3\times64$, lattice spacing $a=0.0934$ fm, and pion mass $m_\pi=260$ MeV. Matrix elements are computed for hadron momenta $P_3=0$, 0.41, 0.83, 1.25, 1.66, and 2.07 GeV and analyzed within the short-distance factorization framework. We investigate the dependence of the extracted moments on the truncation of the operator-product expansion, the coordinate-space fit window, and the perturbative accuracy of the Wilson coefficients, comparing next-to-leading-order and next-to-next-to-leading-order results. We also perform an RG-improved analysis as a consistency check of the perturbative treatment. Our final results are obtained from combined fits in $(P_3,z)$ space at next-to-next-to-leading-order and are quoted at $\mu=2$ GeV. We also study the SU(3) symmetry-breaking effect and reconstruct the valence PDFs from the moments. 
\end{abstract}


\maketitle

\section{Introduction}

Understanding the internal structure of hadrons in terms of quarks and gluons is a central goal of quantum chromodynamics (QCD). Among hadronic systems, the pion and kaon play a particularly important role. As the pseudo--Nambu-Goldstone bosons associated with the spontaneous breaking of chiral symmetry, these mesons provide a unique opportunity to study the interplay between quark masses, chiral dynamics, and the emergence of hadron structure from QCD. In addition, comparing the partonic structure of the pion and kaon provides insight into SU(3) flavor symmetry breaking and the role of the strange quark in shaping the structure of the kaon, with respect to the effect of the up quark.

The collinear partonic structure of hadrons is encoded in parton distribution functions (PDFs), which describe the probability of finding a quark or gluon carrying a fraction $x$ of the hadron's longitudinal momentum. PDFs are essential inputs for theoretical predictions of high-energy scattering processes and provide a fundamental description of hadron structure. While the PDFs of the nucleon are now well constrained by extensive experimental programs, the PDFs of light mesons remain comparatively less well-known due to the scarcity of direct experimental data. Existing experimental information on the pion PDF primarily comes from pion-induced Drell--Yan experiments~\cite{Conway:1989fs,Aicher:2010cb}, while constraints on the kaon structure are even more limited. Consequently, first-principles calculations from lattice QCD play an essential role in improving our knowledge of meson structure.

Lattice QCD provides a first-principles framework for investigating hadron structure, in which correlation functions are computed numerically in Euclidean spacetime, enabling direct access to many nonperturbative observables. 

However, PDFs are defined in terms of light-cone correlations of quark and gluon fields and therefore cannot be computed directly in Euclidean lattice QCD. For many years, lattice studies of partonic structure were consequently restricted to Mellin moments of distribution functions, which correspond to matrix elements of local operators and can be evaluated in Euclidean space~\cite{Martinelli1987,Gockeler1996}. These moments are important observables in their own right, encoding fundamental properties of hadrons such as the fraction of the hadron momentum carried by quarks. Although such calculations have provided valuable information about hadron structure, the determination of higher moments becomes increasingly challenging because of operator mixing, power-divergent renormalization, and rapidly deteriorating signal-to-noise ratios~\cite{Martinelli:1994ty,Gockeler:1998ye}. Due to these limitations, early attempts to reconstruct PDFs from a finite set of Mellin moments were largely inconclusive regarding the feasibility of reliably determining the $x$ dependence (see, e.g.,~\cite{Detmold:2001dv,Holt_2010}). A more recent calculation, however, which benefits from significantly improved statistical and analytical techniques, indicates that it is possible to capture the main features of the PDF using moments up to $\langle x^3 \rangle$~\cite{Alexandrou:2021mmi}.
Lattice calculations of pion and kaon Mellin moments from local operators are progressing from exploratory determinations of low moments and tensor observables to modern calculations with improved control over quark masses, lattice spacings, and renormalization~\cite{Brommel:2006zz,Brommel:2006ww,Brommel:2007zz,Brommel:2007xd,Boyle:2008yd,Kaneko:2008kx,JLQCD:2009ofg,Kaneko:2010ru,Bali:2013gya,Gulpers:2013uca,Fukaya:2014jka,Owen:2015gva,Abdel-Rehim:2015owa,Chambers:2017tuf,Koponen:2017xmw,Alexandrou:2017blh,Wang:2018pii,Oehm:2018jvm,Colangelo:2018mtw,Alexandrou:2020gxs,Wang:2020nbf,Alexandrou:2021mmi,Loffler:2021afv,ExtendedTwistedMass:2021rdx,Alexandrou:2021ztx,Gao:2021xsm,Hackett:2023nkr,ExtendedTwistedMass:2024kjf,Alexandrou:2026nsl}. A more recent approach utilizes gradient flow for the determination of Mellin moments of any order~\cite{Shindler:2023xpd}, and has been implemented for the pion, demonstrating feasibility up to $\langle x^5 \rangle$~\cite{Francis:2025pgf,Francis:2025rya}.

Over the past decade, new theoretical developments have significantly expanded the range of partonic observables accessible in lattice QCD. In particular, large-momentum effective theory (LaMET)~\cite{Ji:2013dva,Ji:2014gla} makes it possible to extract the $x$-dependence of PDFs from Euclidean correlation functions by computing matrix elements of nonlocal quark bilinear operators between hadron states boosted to large momentum. These matrix elements can be related to light-cone PDFs through perturbative matching and large-momentum expansions. A closely related approach is the short-distance factorization~\cite{Radyushkin:2017cyf,Radyushkin:2018cvn} and techniques, which provide complementary ways of accessing partonic distributions from coordinate-space correlation functions.
These methods have enabled a growing number of lattice calculations of the PDFs, including the pion~\cite{Zhang:2018nsy,Izubuchi:2019lyk,Joo:2019bzr,Sufian:2019bol,Sufian:2020vzb,Lin:2020ssv,Gao:2020ito,Gao:2022iex,Holligan:2024umc,Gao:2025inf,Miller:2025wgr}, and limited studies of the kaon~\cite{Lin:2020ssv,Miller:2025wgr}. Such calculations have demonstrated that nonlocal lattice matrix elements can be used to reconstruct the Bjorken-$x$ dependence of meson PDFs, and in some cases, with increasing control over systematic uncertainties. Matrix elements of nonlocal operators also offer an alternative method for determining Mellin moments, which utilizes operator product expansion (OPE). 
This offers a complementary perspective to $x$-space reconstructions and serves as a valuable benchmark for validating PDF reconstructions obtained from nonlocal lattice matrix elements. In this approach, the short-distance behavior of the coordinate-space matrix element of a Wilson-line operator is expanded in powers of the quark separation. The coefficients of this expansion are directly related to the Mellin moments of the PDF. Because the moments are obtained from the same nonlocal matrix elements used in quasi- and pseudo-distribution analyses, this strategy allows one to extract multiple moments without explicitly constructing the corresponding higher-dimensional local operators, thereby alleviating issues such as operator mixing and power-divergent renormalization. This method has been successfully implemented in lattice QCD for both PDFs~\cite{Karpie:2018zaz,Joo:2020spy,Bhat:2020ktg,Gao:2020ito,Fan:2020nzz,Gao:2022iex,Gao:2022uhg,Gao:2023ktu} and GPDs~\cite{Chen:2019lcm,Lin:2023gxz,Bhattacharya:2023ays,Bhattacharya:2024wtg,HadStruc:2024rix,Ding:2024saz,Gao:2025inf}. In a recent work~\cite{Miller:2025wgr}, we presented a lattice QCD determination of the pion and kaon unpolarized PDFs using both the large-momentum effective theory and short-distance factorization frameworks. The present work extends that analysis by focusing on the Mellin moments of the pion and kaon PDFs obtained from the same set of lattice matrix elements.

This paper is organized as follows. In Sec.~\ref{sec:theory}, we review the theoretical framework and define the Mellin moments of the pion and kaon PDFs relevant to our analysis. Section~\ref{sec:lattice_setup} describes the lattice setup and the extraction of the matrix elements entering the calculation. In Sec.~\ref{sec:results}, we present our results for the Mellin moments with different analysis strategies. We also summarize our final results, accounting for systematic uncertainties, and reconstruct the $x$ dependence of the pion and kaon PDFs. In Sec.~\ref{sec:comparison} we provide a comparison with other lattice-QCD studies for the Mellin moments, as well as a comparison with phenomenological estimates for the PDFs. In Sec.~\ref{sec:su3}, we investigate SU(3) flavor-symmetry breaking effects by comparing the moments of the pion and kaon PDFs.  Finally, in Sec .~\ref {sec:conclusions} we summarize our conclusions and outline prospects for future studies.

\section{Theoretical Setup}
\label{sec:theory}

The necessary input for this work is the pion and kaon matrix elements of the nonlocal vector current in the forward limit, which are defined as \\[-3ex]
\begin{equation}
{\mathcal F}_M^f(z, p) \equiv 
\langle M(p) | \bar{\Psi}(0)\gamma_\mu\, {\cal W}(0,z)\Psi(z) | M(p) \rangle \,,
\label{eq:def}
\end{equation}
where $M$ denotes the meson state under consideration. Here, we are interested in the individual quark, $f$, contribution for the pion (up quark) and kaon (up and strange quark). The four-vector momentum of the initial and final states is the same, and such that $\vec{p}\equiv(0,0,P_3)$, which is the required kinematic configuration for accessing the pion and kaon parton distribution functions from nonlocal operators. The operator entering the matrix element consists of quark fields separated along the spatial direction $\hat z$ and connected by a straight Wilson line ${\cal W}(0,z)$ to ensure gauge invariance. We employ the Dirac structure $\gamma^0$, while the quark separation is taken along the boost direction. This choice is motivated by the structure of the light-cone correlation function, where the operator $\gamma^+$ appears naturally. The $\gamma^3$ component of  $\gamma^+$ is not ideal as it mixes with the scalar operator under renormalization~\cite{Constantinou:2017sej,Alexandrou:2017huk}, due to chiral symmetry breaking. This mixing is a lattice artifact and disappears in the continuum limit. In the case of the twisted-mass formulation, which is used in this work, the mixing occurs between $\gamma_3$ and $\gamma_5$ in the twisted basis. Since the forward matrix element of the $\gamma_5$ operator vanishes, the resulting effect manifests only as additional gauge noise, as demonstrated explicitly for the nucleon in Ref.~\cite{Alexandrou:2019lfo}. To avoid such complications and the increased noise, we restrict ourselves to the $\gamma_0$ operator, which is multiplicatively renormalizable, a particularly important property when using ratios of these matrix elements (Eq.~\eqref{eqn:DR}).

For a given quark flavor $f$, the resulting ground-state matrix element $F_M^f$ is related to the corresponding unpolarized light-cone PDF $q_M^f$. In this work, we focus on the Mellin moments of the PDFs, which we extract using the procedure described here. First, to remove divergence associated with the Wilson-line operator, we construct the reduced Ioffe-time distribution through the double ratio
\begin{equation}
    \mathcal{M}(\nu,z^2) = \frac{F_M^f(\nu,z^2)/F_M^f(\nu,0)}{F_M^f(0,z^2)/F_M^f(0,0)}\,,
    \label{eqn:DR}
\end{equation}
where $\nu \equiv z P_3$ is the Ioffe time. By construction, this ratio cancels the leading ultraviolet divergence of the Wilson line and normalizes the matrix element such that $\mathcal{M}(0,z^2)=1$. The use of the reduced ITD is particularly convenient for a short-distance analysis. At sufficiently small $z^2$, the reduced ITD can be analyzed using the OPE. In this regime, the nonlocal operator is expanded in a tower of local twist-two operators, and the reduced ITD can be written as 
\begin{equation}
\label{eq:expansion}
\mathcal{M}(\nu,z^2) = \sum_{n=0}^{\infty} 
\frac{(i\nu)^n}{n!}\, c_n(z^2\mu^2,\alpha_s)\, \langle x^n\rangle \,,
\end{equation}
where $\langle x^n\rangle$ are the Mellin moments of the corresponding light-cone PDF and $c_n$ are perturbatively calculable Wilson coefficients. Thus, the short-distance behavior of the nonlocal matrix element is fully determined by the Mellin moments, up to higher-twist corrections and truncation effects in the expansion. At next-to-leading order (NLO), the Wilson coefficients can be computed from the matching kernel of the pseudo-distribution formalism (see Appendix~\ref{appA}), leading to the explicit expression
\begin{equation}
    \mathcal{M}(\nu,z^2) = \sum_{n=0}^{\infty}\frac{(i\nu)^n}{n!} \langle x^n\rangle\left\{1 {+} \frac{\alpha_sC_F}{2\pi}\left[\ln\left(z^2\mu^2\frac{e^{2\gamma_E{+}1}}{4}\right)\left(H_n {+} H_{n{+}2} {-} \frac{3}{2}\right) {+} 2\left(\frac{1}{n{+}1}{-}\frac{1}{n{+}2}{-}H_n^2{-}H_n^{(2)}\right){-}1\right]\right\}\,,
    \label{eqn:moment_expansion}
\end{equation}
where $H_n=\sum_{k=1}^n \frac{1}{k}$ and $H_n^{(2)}=\sum_{k=1}^n \frac{1}{k^2}$ are harmonic numbers. In this work, we employ Wilson coefficients computed at both NLO and NNLO accuracy (see, e.g., Refs.~\cite{Bhat:2022zrw,PhysRevLett.126.072001}), which allows us to assess the impact of higher-order perturbative corrections on the extracted moments.

It is useful to clarify how the even and odd Mellin moments are accessed from the reduced Ioffe-time distribution. We use the standard convention in which the PDF is defined on the interval $x\in[-1,1]$, with the negative-$x$ region related to the antiquark distribution by $q(-x)=-\bar q(x)$. The light-cone Ioffe-time distribution may be written as
\begin{equation}
    \mathcal{Q}(\nu,\mu)=\int_{-1}^{1} dx\, e^{i\nu x} q(x,\mu)
    =
    \int_0^1 dx\,\cos(\nu x)\,[q(x,\mu)-\bar q(x,\mu)]
    + i\int_0^1 dx\,\sin(\nu x)\,[q(x,\mu)+\bar q(x,\mu)] \,.
\end{equation}
Thus, the real part is sensitive to the $q-\bar q$ combination, while the imaginary part is sensitive to the $q+\bar q$ combination.
Equivalently, the Mellin moments of the distribution defined on $[-1,1]$,
\begin{equation}
\langle x^n\rangle \equiv \int_{-1}^{1} dx\, x^n q(x,\mu)\,,
\end{equation}
can be written as
\begin{equation}
\langle x^n\rangle
=
\int_0^1 dx\, x^n
\left[q(x,\mu)+(-1)^{n+1}\bar q(x,\mu)\right]\,.
\end{equation}
Therefore, even moments correspond to the valence combination,
\begin{equation}
\langle x^{2m}\rangle
=
\int_0^1 dx\, x^{2m}\,[q(x,\mu)-\bar q(x,\mu)]\, ,
\end{equation}
whereas odd moments correspond to the $q+\bar q$ combination,
\begin{equation}
\langle x^{2m+1}\rangle
=
\int_0^1 dx\, x^{2m+1}\,[q(x,\mu)+\bar q(x,\mu)] \,.
\end{equation}
In the analysis below, we use the notation $\langle x^n\rangle$ throughout, with the understanding that the even moments are obtained from the real part of the reduced Ioffe-time distribution and the odd moments from the imaginary part.

The expansion of Eq.~\eqref{eqn:moment_expansion} makes explicit that the dependence of the reduced Ioffe-time distribution on $\nu$ is governed by the order, $n$, of Mellin moments $\langle x^n\rangle$, while the $z^2$ dependence enters through perturbatively calculable logarithmic and finite terms. This expression forms the basis for extracting moments from the lattice data. In practice, the infinite series in Eq.~\eqref{eqn:moment_expansion} is truncated at a finite order $n_{max}$, and the Mellin moments are determined from fits to the lattice data for $\mathcal{M}(\nu,z^2)$. In this work, we consider $n_{max}=2,\cdots,6$, implemented within two complementary fitting strategies: input data on a single $z$ value (Sec.~\ref{sub:z_fixed}), or span over a range of $z$ values (Sec.~\ref{sub:P3_z_combined}). Systematic uncertainties may be estimated by varying $n_{max}$ and the fit in $z$. More details on these fits are presented in Sec.~\ref{sub:z_fixed} - \ref{sub:P3_z_combined}. 

Another element of the calculation regards the renormalization scale, and here we use two approaches. In the first one, the perturbative matching is performed directly at a fixed renormalization scale, such that the moments are obtained at the target scale without an explicit evolution step (Sec.~\ref{sub:fixed}). In the second approach, matching is carried out at an intermediate scale, and the resulting moments are then evolved to the reference scale via perturbative renormalization-group evolution (Sec.~\ref{sub:evol}). These two procedures provide a useful check on the stability of the extraction and allow us to disentangle effects associated with the short-distance expansion from those associated with perturbative evolution.

\section{Computational Setup}
\label{sec:lattice_setup}

The results presented in this work are obtained from the calculation of Ref.~\cite{Miller:2025wgr}, which uses gauge configurations generated by the Extended Twisted Mass Collaboration (ETMC)~\cite{Alexandrou:2018egz,ExtendedTwistedMass:2021gbo}. The simulations employ twisted-mass fermions with a clover term and the Iwasaki-improved gauge action. The ensemble (\texttt{cA211.30.32}) includes two degenerate light quarks and dynamical strange and charm quarks in the sea ($N_f=2+1+1$).  The pion mass is approximately $260$ MeV on a lattice of size $32^3\times 64$ with spacing $a\simeq0.093$ fm. Further details regarding the ensemble generation and quark-mass tuning can be found in Refs.~\cite{Alexandrou:2018egz,ExtendedTwistedMass:2021gbo}. The main parameters of the ensemble are summarized in Table~\ref{tab:params}.

\begin{table}[h!]
\centering
\renewcommand{\arraystretch}{1.4}
\renewcommand{\tabcolsep}{6pt}
\begin{tabular}{| l| c | c | c | c | c  | c | c |}
\hline
\multicolumn{8}{|c|}{Parameters} \\
\hline
Ensemble & $\beta$ & $a$ [fm] & volume $L^3\times T$ & $N_f$ & $m_\pi$ [MeV] &
$L m_\pi$ & $L$ [fm]\\
\hline
cA211.30.32 & 1.726 & 0.0934 & $32^3\times 64$ & 2+1+1 & 260 & 4 & 3.0 \\
\hline
\end{tabular}
\caption{\small Parameters of the ensemble used in this work.}
\label{tab:params}
\end{table}

The pion and kaon matrix elements are obtained from correlation functions constructed with boosted meson states. For a meson $M$ carrying spatial momentum $\vec p$, the two-point function is
\begin{equation}
C^{\rm 2pt}_M(\vec p,t)=
\sum_{\vec x} e^{-i\vec p\cdot \vec x}
\langle J_M(t,\vec x) J^\dagger_M(0,0)\rangle \,.
\end{equation}
To access the nonlocal matrix elements entering Eq.~\eqref{eq:def}, we compute the corresponding three-point functions
\begin{equation}
\label{eq:3pt}
{\mathcal C}_M^f(\vec p;z,t_s,t)=
\sum_{\vec x_s,\vec x}
e^{-i\vec p\cdot\vec x_s}
\langle J_M(t_s,\vec x_s)
{\mathcal O}^f_{\gamma_0}(t,\vec x;z)
J^\dagger_M(0,0)\rangle \,,
\end{equation}
where $t$ and $t_s$ denote the insertion and sink Euclidean times, respectively. Throughout this work, we consider the forward kinematics in which the initial and final meson states carry the same momentum. Focusing on the connected diagram, the three-point functions are evaluated via sequential inversion through the sink, enabling us to compute matrix elements of several operators with minimal computational overhead.
The matrix elements of Eq.~\eqref{eq:def} are obtained from the ratio
\begin{equation}
R_M^f(z,p;t_s,t)=
\frac{{\mathcal C}_M^f(z,p;t_s,t)}
{C_M^{2pt}(p;t_s)}\,\, \xrightarrow[t \gg a]{t_s-t \gg a} \, F_M^f(z,p)\,,
\label{eqn:plateau}
\end{equation}
and the ground-state matrix element is identified from the plateau region
of the ratio to suppress excited-state effects. We utilize a fit range of $t \in [3a-9a]$ when $t_s = 12a$ and $t\in [3a-7a]$ when $t_s = 10a$ on both the real and imaginary parts of $F_{\pi}^u,~F_{K}^u$ and $F_K^s$. More details about the lattice calculation can be found in Ref.~\cite{Miller:2025wgr}.

The extraction of partonic observables requires boosting the meson along the $z$ direction as seen in Eq.~\eqref{eq:3pt}. Here, we calculated the matric elements at six values of the momentum boost, that is,
\begin{equation}
    P_3 = 0,\;\pm0.41,\;\pm0.83,\;\pm1.25,\;\pm1.66,\;\pm2.07~{\rm GeV}\,.
\end{equation}
As the boost increases, the signal-to-noise ratio deteriorates rapidly. To enhance the overlap with the boosted ground state, we apply momentum smearing~\cite{Bali:2016lva} to the quark fields. The gauge links in the nonlocal operators are stout-smeared with five iterations and a parameter $\rho=0.15$ to reduce gauge noise. To maintain statistical precision, we adjust both the source–sink separation and the number of source positions per configuration. For the pion, momenta below $1$ GeV are computed at $t_s=12a$, while larger boosts use $t_s=10a$. For the kaon, boosts up to $P_3=1.25$ GeV are evaluated at $t_s=12a$. The number of source positions is increased significantly at large
momenta, reaching up to 400 (300) sources per configuration for the
pion (kaon) at the highest boost.
The resulting statistics are listed in Table~\ref{tab:stats}.
\begin{table}[h!]
\begin{center}
\renewcommand{\arraystretch}{1.5}
\begin{tabular}{l|cccccc}
\hline
$P_3$ [GeV] & $\quad$0$\quad$ & $\quad$$\pm$0.41$\quad$ & $\quad$$\pm$0.83$\quad$ & $\quad$$\pm$1.25$\quad$ & $\quad$$\pm$1.66$\quad$ & $\quad$$\pm$2.07$\quad$  \\ \hline
$t_s/a$ & 12 & 12 & 12 & 10$^\star$, 12$^\dagger$ & 10 & 10 \\\hline
$N_{\rm confs}$ & 1,198 & 1,198 & 1,198 & 1,198 & 1,198 & 1,198 \\\hline
$N^{\pi}_{\rm src}$ & 1 &  8 &  8 &  56 &  84 & 400  \\\hline
$N^{K}_{\rm src}$ & 1 &  8 &  8 &  8 &  48 & 300  \\\hline
$N^{\pi}_{\rm tot}$ & 1,198 & 9,584 & 9,584 & 67,088  & 100,632  & 479,200 \\\hline
$N^{K}_{\rm tot}$ & 1,198 & 9,584 & 9,584 & 9,584  & 57,504  & 359,400 \\
\hline
\end{tabular}
\caption{\small Statistics for the pion and kaon matrix elements at different values of momentum $P_3$. $N_{\rm confs}$, $N^{\pi}_{\rm src}$, $N^K_{\rm src}$, $N^{\pi}_{\rm total}$ and $N^K_{\rm total}$ denote the number of configurations, source positions per configuration, and total statistics, respectively. \\[0.75ex] $\star$: pion; $\dagger$: kaon.}
\label{tab:stats}
\end{center}
\end{table}

\section{Lattice Results}
\label{sec:results}

\begin{figure}[h!]
\hspace*{-0.2cm}    \includegraphics[scale=0.29]{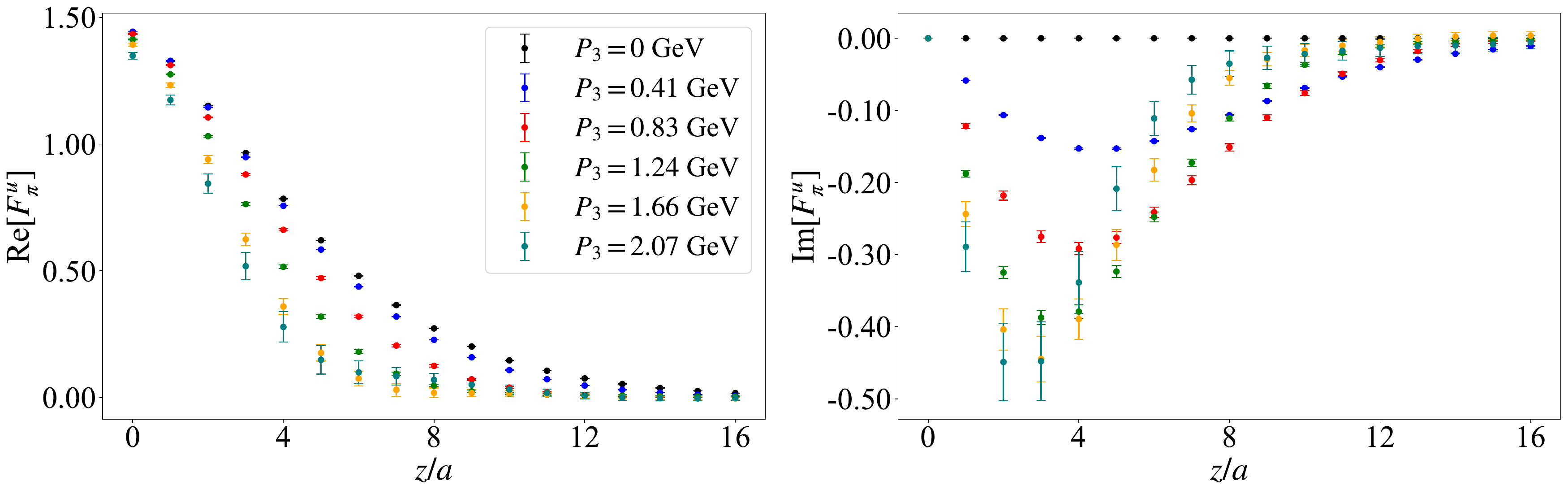}
    \vspace*{-0.35cm}
    \caption{\small{Bare matrix element $F_\pi^u$ for momentum boost $|P_3|=0,~0.41,~0.83,~1.25,~1.66,~2.07$ GeV. }}
    \label{fig:g0_pion}
\end{figure}

\begin{figure}[h!]
\hspace*{-0.2cm}   \includegraphics[scale=0.29]{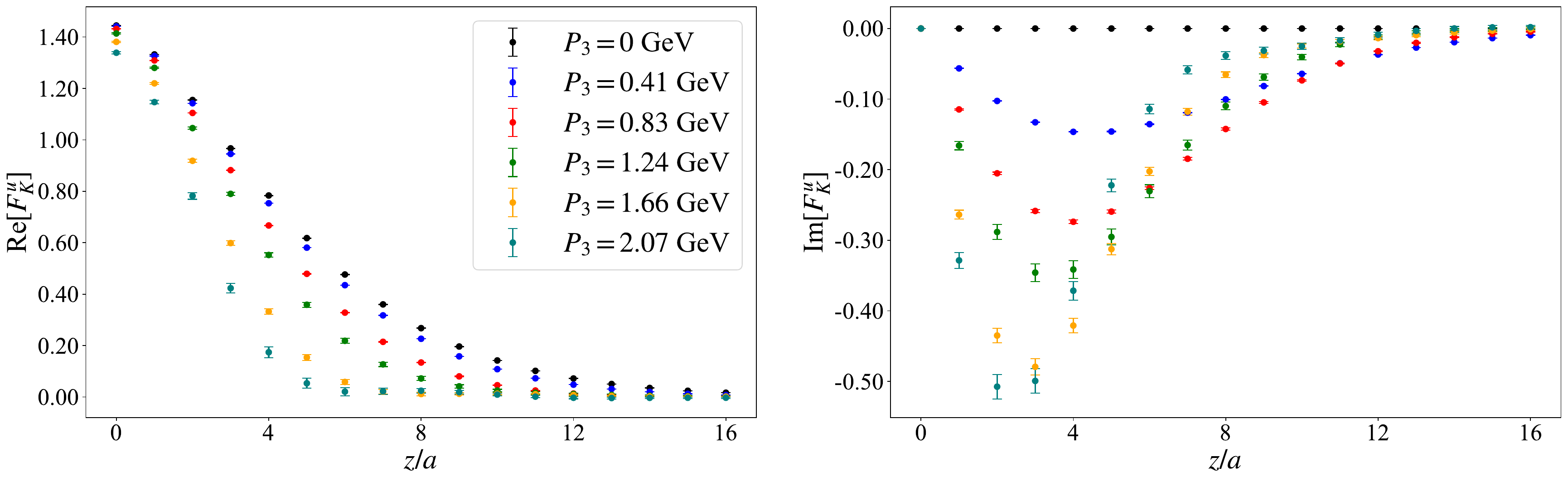}
        \vspace*{-0.35cm}
    \caption{\small{Bare matrix elements $F_K^u$. The notation is the same as Fig.~\ref{fig:g0_pion}.}}
    \label{fig:g0_kaon_u}
\end{figure}

\begin{figure}[h!]
\hspace*{-0.2cm}    \includegraphics[scale=0.29]{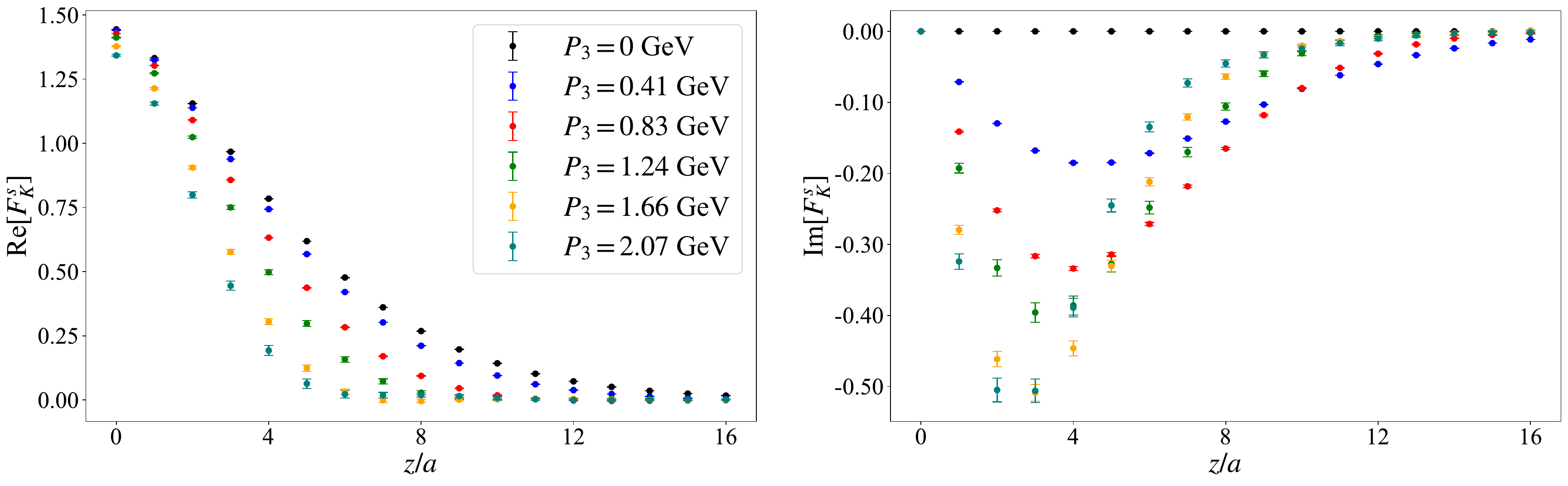}
    \vspace*{-0.35cm}
    \caption{\small{Bare matrix elements $F_K^s$. The notation is the same as Fig.~\ref{fig:g0_pion}.}}
    \label{fig:g0_kaon_s}
\end{figure}

\begin{figure}[!h]
    \centering
    \includegraphics[scale=0.29]{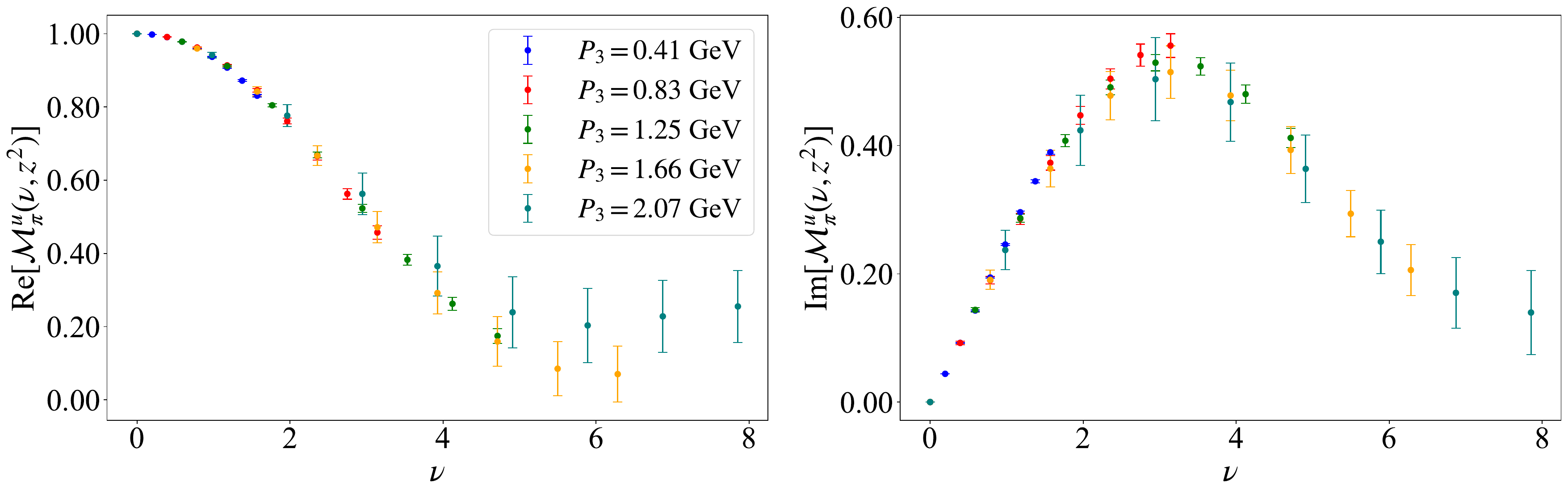}
        \vspace*{-0.35cm}
    \caption{Double ratio $\mathcal{M}_\pi^u$ for $|P_3| \in \{0.41,0.83,1.25,1.66,2.07\}$ GeV for the real (left) and imaginary (right) parts.}
    \label{fig:DR_pion}
\end{figure}

\begin{figure}[!h]
    \centering
    \includegraphics[scale=0.29]{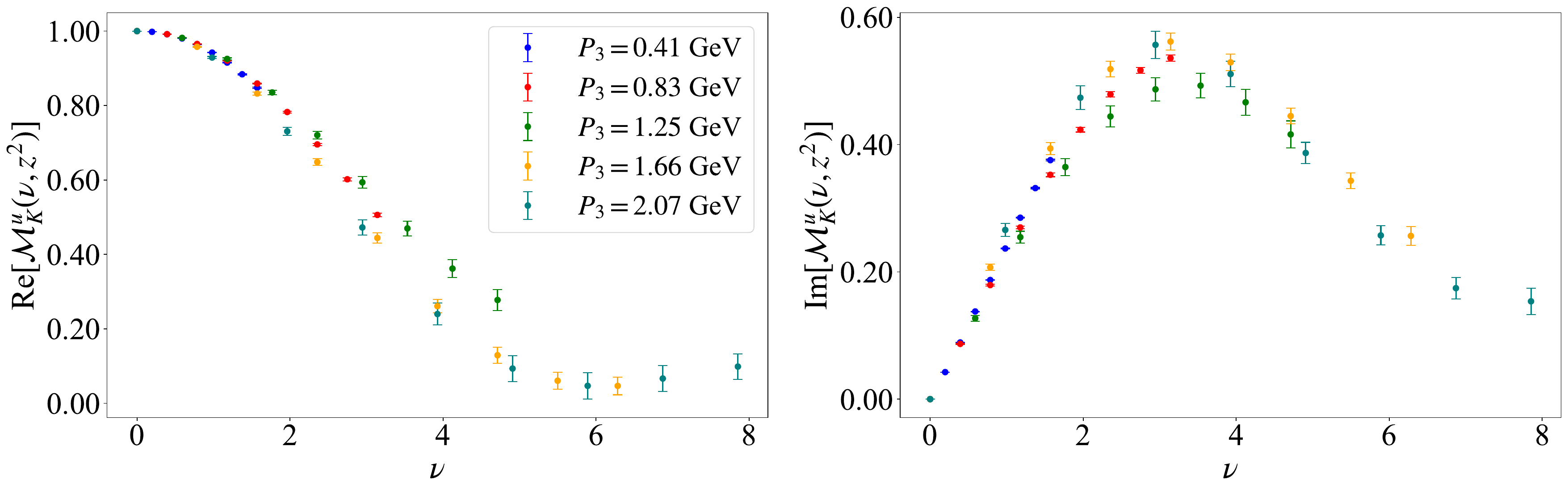}
        \vspace*{-0.35cm}
    \caption{Double ratio $\mathcal{M}_K^u$. The notation is the same as Fig.~\ref{fig:DR_pion}.}
    \label{fig:DR_kaon_u}
\end{figure}

\begin{figure}[!h]
    \centering
    \includegraphics[scale=0.29]{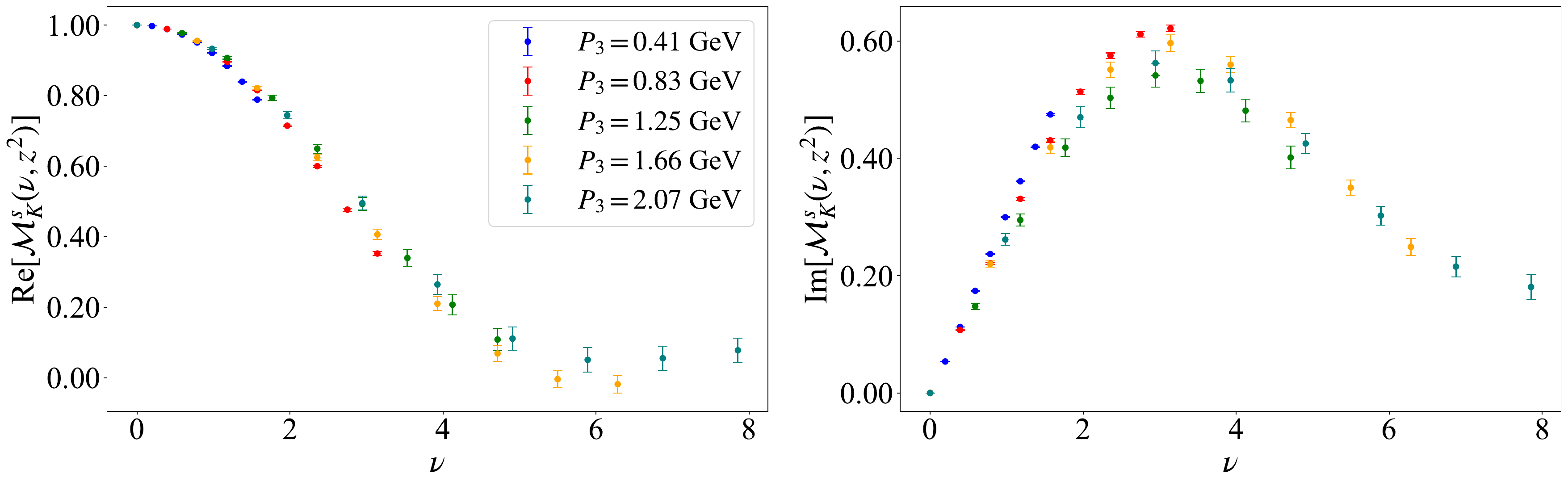}
    \caption{Double ratio $\mathcal{M}_K^s$. The notation is the same as Fig.~\ref{fig:DR_pion}.}
    \label{fig:DR_kaon_s}
\end{figure}

The first step of the analysis is the extraction of the ground-state matrix elements defined in Eq.~\eqref{eq:def}, from which one forms the double ratio of Eq.~\eqref{eqn:DR}. These have been presented and discussed in Ref.~\cite{Miller:2025wgr}, but we also show the matrix elements in
Figs.~\ref{fig:g0_pion} - \ref{fig:g0_kaon_s}, and the double ratios in Figs.~\ref{fig:DR_pion} - \ref{fig:DR_kaon_s}, for completeness and ease of reference. 
In the aforementioned plots, we have explicitly imposed the symmetry relation $F_M^f(-z\cdot P) = F_M^{f*}(z\cdot P)$ to improve statistical precision. From the plots, it can be seen that as the boost increases, the statistical noise increases, despite the significant increase in the statistics for $P_3 \geq 1.25$ GeV for the pion, and for $P_3 \geq 1.66$ GeV for the kaon. Even at the current statistics, we find that $P_3 = 1.65$ and $2.07$ GeV have lower signal-to-noise ratios than smaller values of the momentum boost.

\newpage

\subsection{Fixed-order Mellin moments}
\label{sub:fixed}

We first present the results for the Mellin moments obtained by setting the scale in the matching expressions (e.g., Eq.~\eqref{eqn:moment_expansion}) equal to the desired scale, which, in this work, is chosen to be $\mu=2~\mathrm{GeV}$. In this setup, the Wilson coefficients entering the short-distance expansion are evaluated at the above scale and are used to connect the lattice matrix elements to the Mellin moments without introducing an additional evolution step. This procedure allows us to examine the dependence of the extracted moments on the expansion truncation and the choice of fit range in coordinate space. In particular, one may assess how well the lattice data constrain the lowest moments before incorporating perturbative scale evolution.

\subsubsection{Moments at fixed $z$ value}
\label{sub:z_fixed}

A possible method for extracting the Mellin moments is to perform fits at fixed $z$, following the procedure of Refs.~\cite{Gao:2022iex,Su_2023,Bhattacharya:2023ays,Bhattacharya:2024wtg}. In this approach, the reduced Ioffe-time distribution $\mathcal{M}(\nu,z^2)$ is analyzed independently at each value of $z$ by fitting its $\nu$-dependence using the truncated expansion of Eq.~\eqref{eqn:moment_expansion}. For a given value of $z$, the Mellin moments are obtained by minimizing a $\chi^2$ function constructed over the available momenta $P_3$, that is,
\begin{equation}
    \chi^2 = \sum_{P_3} \frac{(\mathrm{Re/Im}[\mathcal{M}^{\rm SDF}(\nu,z^2)] - \mathrm{Re/Im}[\mathcal{M}^{\rm data}(\nu,z^2)])^2}{\sigma^2_{\mathrm{Re/Im}}}\,,
    \label{eqn:chi2_fixed_z}
\end{equation}
where the real and imaginary parts are fitted separately for each $z$. In contrast to the combined $(P_3,z)$ fit examined in Sec.~\ref{sub:P3_z_combined} below, this procedure does not correlate data across different $z$ values, and therefore yields a set of Mellin moments that depend on such a choice. As a result, one may identify the residual $z$-dependence. 

\begin{figure}[!h]
    \centering
\includegraphics[scale=0.242]{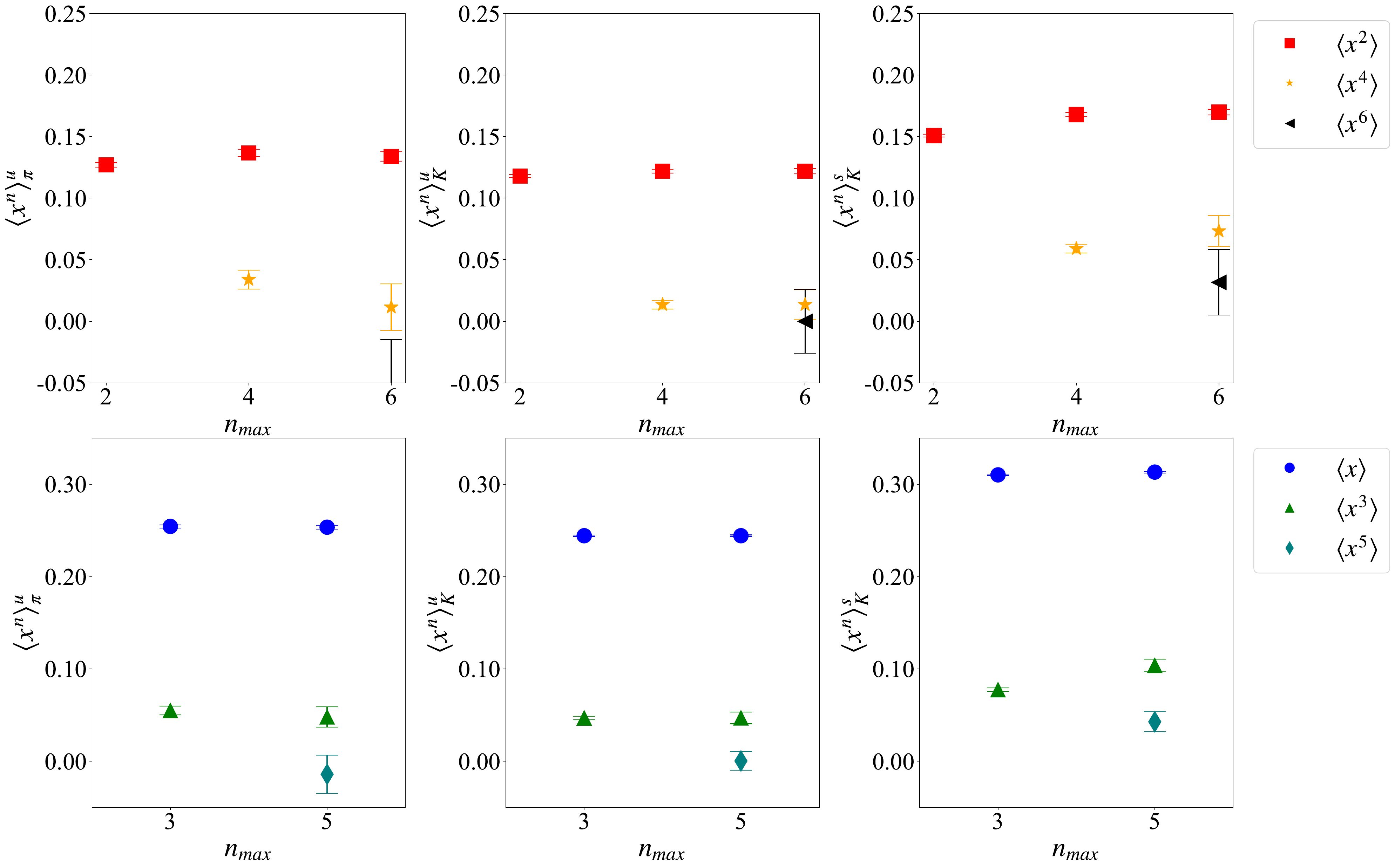}
\vspace*{-0.2cm}
    \caption{Top: Even-order Mellin moments obtained at $n_{max}= 2,4,6$ at fixed $z = 4a$. Bottom: Odd-order Mellin moments obtained at $n_{max}=3,5$ at fixed $z = 4a$. NLO accuracy has been used for the Wilson coefficients. }
\label{fig:fixed_z4_nmax_test}
\end{figure}

While Eq.~\eqref{eqn:moment_expansion} is an infinite sum, we must truncate at some $n_{max}$. Even (odd) moments are linked to the real (imaginary) part of the fitted ratio. Here, we tested $n_{max} = 2, 4, 6$ for the real part and $n_{max}= 3, 5$ for the imaginary part. In Fig.~\ref{fig:fixed_z4_nmax_test}, we show the extracted moments for both the pion and the kaon at $z=4a$. These data include NLO accuracy for the Wilson coefficients. By construction, the choice of $n_{max}$ determines the highest Mellin moment that can be extracted from the truncated expansion. For example, selecting $n_{max}=2$ allows access to moments up to $\langle x^2 \rangle$, while $n_{max}=4$ enables the determination of both $\langle x^2 \rangle$ and $\langle x^4 \rangle$. An analogous procedure applies to extracting the odd moments. It is interesting to note that for all truncations of $n_{max}$, the first two moments listed maintain stability for both the pion and kaon. Also, $\langle x^3 \rangle$ and $\langle x^4 \rangle$ are not sensitive to $n_{max}$ for the up-quark contribution to the pion and kaon. However, for the kaon strange component, $\langle x^3 \rangle_K^s$ and $\langle x^4 \rangle_K^s$ are influenced by systematic uncertainties. Finally, the cases of $\langle x^5 \rangle$ and $\langle x^6 \rangle$ have a non-vanishing signal only for the strange quark contribution to the kaon. Based on the overall behavior, we use $n_{max} = 5$ and $n_{max} =6$ for the odd and even moments, respectively.

\begin{figure}[!h]
    \centering
\includegraphics[scale=0.249]{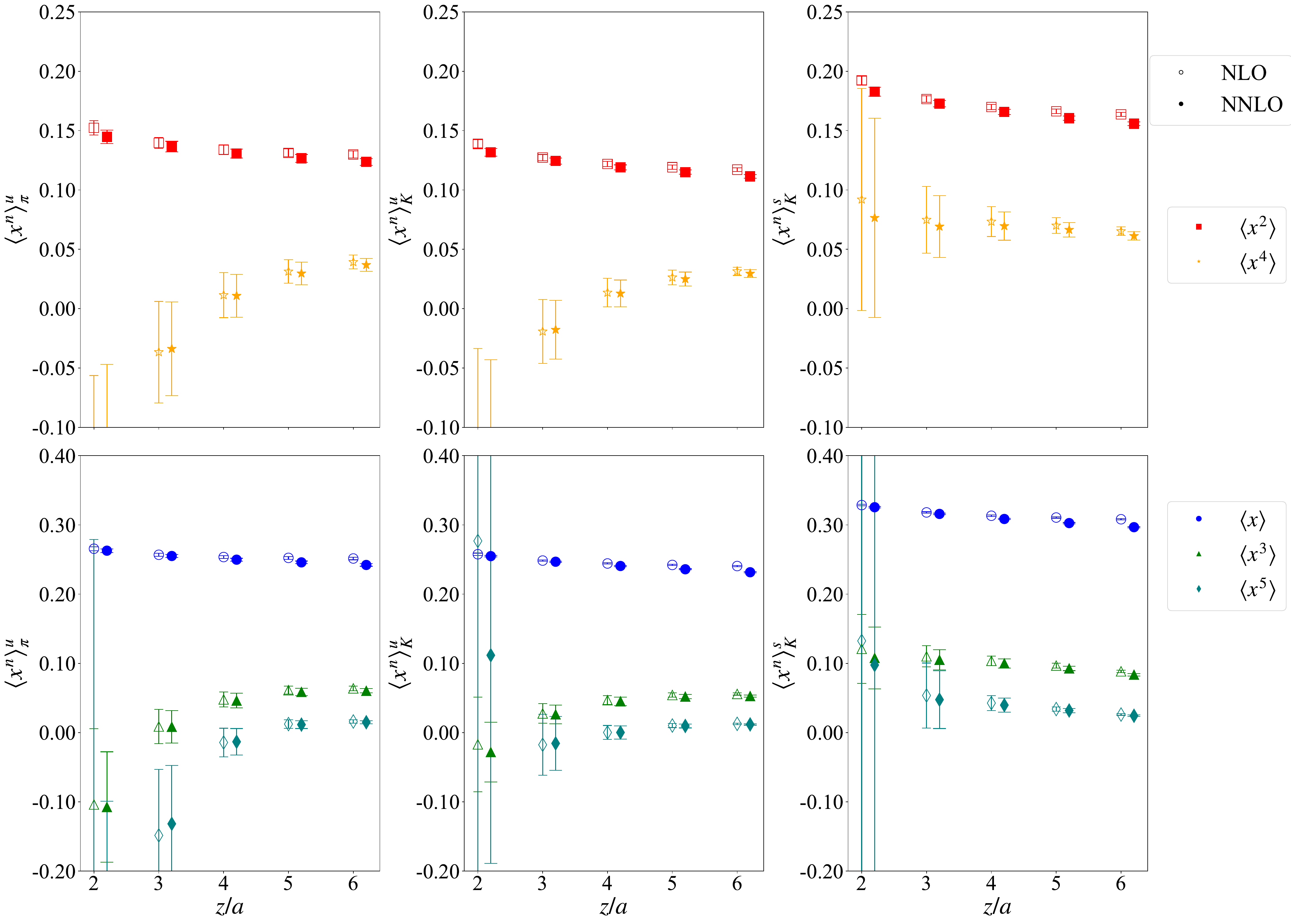}
    \vspace*{-0.25cm}
    \caption{Mellin moments extracted at fixed $z$ (Eq.~\eqref{eqn:chi2_fixed_z}) for the pion (left), kaon up-quark (middle), and kaon strange-quark (right) at NLO (open) and NNLO (filled) accuracy. The top row shows even moments $\langle x^2 \rangle$, $\langle x^4 \rangle$, while the bottom row corresponds
    to the odd moments $\langle x \rangle$, $\langle x^3 \rangle$, and $\langle x^5\rangle$. Results are shown in the $\MSb$ at 2 GeV.}
    \label{fig:fixed_z_moments_NLO_vs_NNLO}
\end{figure}

To assess the $z$ dependence in the Mellin moments, we show $\langle x \rangle$ - $\langle x^6 \rangle$ in Fig.~\ref{fig:fixed_z_moments_NLO_vs_NNLO} as a function of the $z$ value used in the fit. We use $n_{max}=6$ for the even moments and $n_{max}=5$ for the odd ones. The results show that the extracted Mellin moments depend on the choice of $z$ in both the pion and kaon cases, and this dependence becomes more pronounced for higher moments. For the leading moments, $\langle x \rangle$ and $\langle x^2 \rangle$, the extracted values have a relatively mild variation with $z$, particularly for the pion and the up-quark contribution to the kaon. In contrast, the higher moments show substantially larger fluctuations and statistical uncertainties, indicating greater sensitivity to truncation effects in the OPE and to higher-twist contributions. The moments $\langle x^5 \rangle$ and $\langle x^6 \rangle$ are generally poorly constrained, with uncertainties growing rapidly at small values of $z$, where the sensitivity to higher-order terms in the expansion is reduced due to the limited range in Ioffe time. 
The observed $z$ dependence may also indicate competing systematic effects between discretization effects, residual non-perturbative contributions, and higher-twist corrections. The absence of a clear plateau in the moments suggests that a fit based on a single value of $z$ is insufficient to fully isolate these effects. 
Another feature visible in Fig.~\ref{fig:fixed_z_moments_NLO_vs_NNLO} is that the strange-quark moments of the kaon are systematically larger than the corresponding up-quark moments. This is consistent with the physical expectation that the heavier strange quark carries, on average, a larger contribution, as well as observations in other studies~\cite{Miller:2025wgr,Alexandrou:2020gxs,Alexandrou:2021mmi}.

Another interesting investigation concerns the perturbative expansion of the Wilson coefficients. Thus, we compare the moments obtained with the coefficients at NLO and NNLO accuracy, both of which are available in the literature (see, e.g., Refs.~\cite{Bhat:2022zrw,PhysRevLett.126.072001}). This comparison allows us to examine whether the perturbative expansion is sufficiently convergent within the range of Wilson-line separations used in the analysis presented in this Section. In Fig.~\ref{fig:fixed_z_moments_NLO_vs_NNLO} we show the moments extracted using NLO and NNLO Wilson coefficients. We emphasize that the determinations of $\langle x^i \rangle^{\rm NLO}$ and $\langle x^i \rangle^{\rm NNLO}$ are performed independently within the multi-moment fit with $n_{max}=6$. Specifically, Eq.~\eqref{eq:expansion} is fitted separately using either the NLO or NNLO Wilson coefficients, $c_i^{\rm NLO}$ and $c_i^{\rm NNLO}$. The impact of the NNLO coefficients is most visible in the two lowest moments, especially at $z=2a$. Moments with $n=3 - 5$ exhibit negligible dependence within statistical uncertainties. In the comparison of Fig.~\ref{fig:fixed_z_moments_NLO_vs_NNLO}, we exclude $\langle x^6 \rangle$, as it is zero within uncertainties for most of the cases.

Overall, the fixed-$z$ analysis provides valuable insight into the systematic effects arising from extracting Mellin moments from the short-distance expansion. As seen in Fig.~\ref{fig:fixed_z_moments_NLO_vs_NNLO}, the lowest moments exhibit a systematic downward trend with increasing $z$, and the higher moments an upward trend and carry significantly larger uncertainties. The absence of a clear plateau as a function of $z$ indicates that residual higher-twist effects, discretization artifacts, and limitations of the truncated OPE all play a non-negligible role in the extraction. This motivates the combined $(P_3,z)$ analysis presented below, which incorporates information from multiple separations simultaneously.

\subsubsection{Combined $P_3$ and $z$ fits}
\label{sub:P3_z_combined}

\begin{figure}[!h]
    \centering
    \includegraphics[scale=0.26]{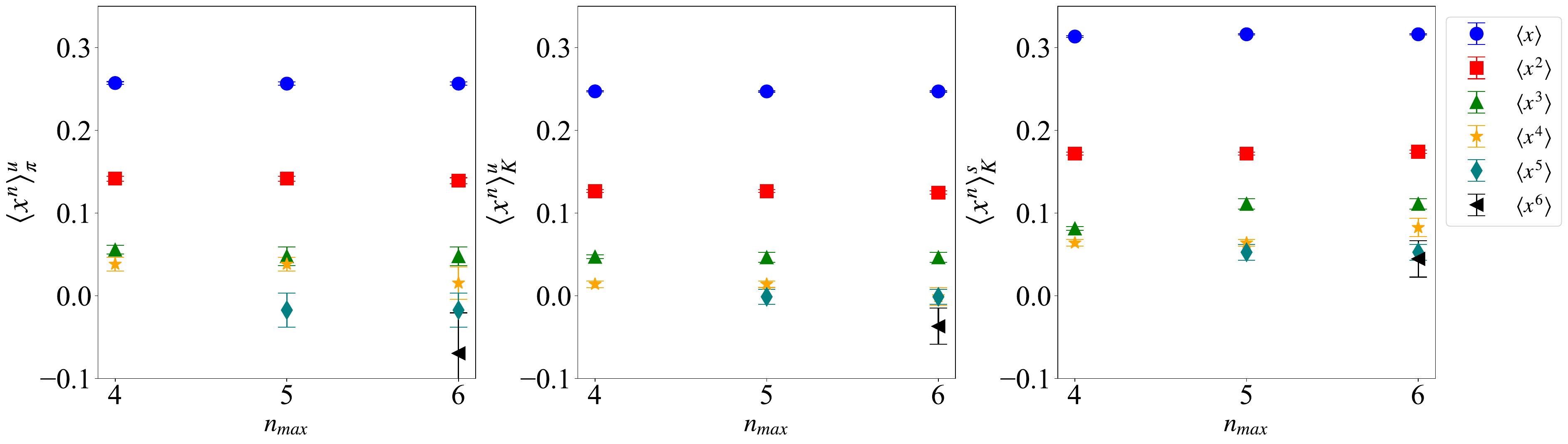}
    \vspace*{-0.5cm}
    \caption{Mellin moments $\langle x^n \rangle^u_\pi$ (left), $\langle x^n \rangle^u_K$ (middle), and $\langle x^n \rangle^s_K$(right) for $n_{max}=4,5,6$  in Eq.~\eqref{eq:expansion}. Results are obtained at NLO accuracy, and the fit includes data for $z \in [3a-4a]$.}
    \label{fig:moments_nmax_test}
\end{figure}

An alternative strategy for extracting the Mellin moments is to perform a global fit to the reduced Ioffe-time distribution using the truncated expansion of Eq.~\eqref{eqn:moment_expansion}. In practice, this requires determining the set of moments $\{\langle x^n\rangle\}$ that best describe the data within a chosen range of $z$ and $P_3$ values, where the OPE is expected to be valid, balancing discretization effects at small $z$ and higher-twist contributions at larger $z$.
To this end, we construct a global $\chi^2$ function that incorporates data points for both the real and imaginary parts of the reduced Ioffe-time distribution, which are described by the same set of Mellin moments. The $\chi^2$ function is defined as
\begin{equation}
    \chi^2 = \sum_{P_3} \sum_{z=z_{min}}^{z_{max}} \left(\frac{(\mathrm{Re}[\mathcal{M}^{\rm SDF}(\nu,z^2)] - \mathrm{Re}[\mathcal{M}^{\rm data}(\nu,z^2)])^2}{\sigma_{\mathrm{Re}}^2}  + \frac{(\mathrm{Im}[\mathcal{M}^{\rm SDF}(\nu,z^2)] - \mathrm{Im}[\mathcal{M}^{\rm data}(\nu,z^2)])^2}{\sigma_{\mathrm{Im}}^2}\right)\,,
    \label{eqn:chi2}
\end{equation}
where $\mathcal{M}^{\rm SDF}$ denotes the truncated OPE expression of Eq.~\eqref{eqn:moment_expansion}, and $\mathcal{M}^{\rm data}$ corresponds to the lattice data obtained from the double ratio of Eq.~\eqref{eqn:DR}. The uncertainties $\sigma_{\mathrm{Re}}$ and $\sigma_{\mathrm{Im}}$ are taken from the statistical errors of the lattice data. The sums in Eq.~\eqref{eqn:chi2} define a global fit over the range of $(P_3,z)$. This increases the number of degrees of freedom relative to the fixed-$z$ fits discussed above and allows the Mellin moments to be constrained simultaneously across multiple boosts and separations $z$. Furthermore, fitting the real and imaginary parts simultaneously enhances sensitivity to both even and odd moments, as these components are governed by different powers in the expansion of Eq.~\eqref{eqn:moment_expansion}. 
We note that, although Ioffe time is the natural variable appearing in the expansion, lattice data exhibit a residual dependence on $P_3$ and $z$ separately (see, e.g., Figs.~\ref{fig:DR_pion} - \ref{fig:DR_kaon_s}). That is, data points corresponding to identical values of $\nu$ but different combinations of $(P_3,z)$ are not statistically equivalent, which is why we keep the variables $z$ and $P_3$ separately. For completeness, we also examine the dependence of the extracted moments on the truncation order $n_{max}$ in the combined $(P_3,z)$ analysis. As discussed in Sec.~\ref{sub:fixed}, increasing $n_{max}$ gives access to higher moments, and, consequently, increases the number of fit parameters entering the analysis. The results are shown in Fig.~\ref{fig:moments_nmax_test}, where we observe that the extracted moments become stable for sufficiently large truncation order, and throughout this work we adopt $n_{max}=6$, which provides stability in the fit and sensitivity to higher-order moments.

\begin{figure}[!h]
    \centering
    \includegraphics[scale=0.23]{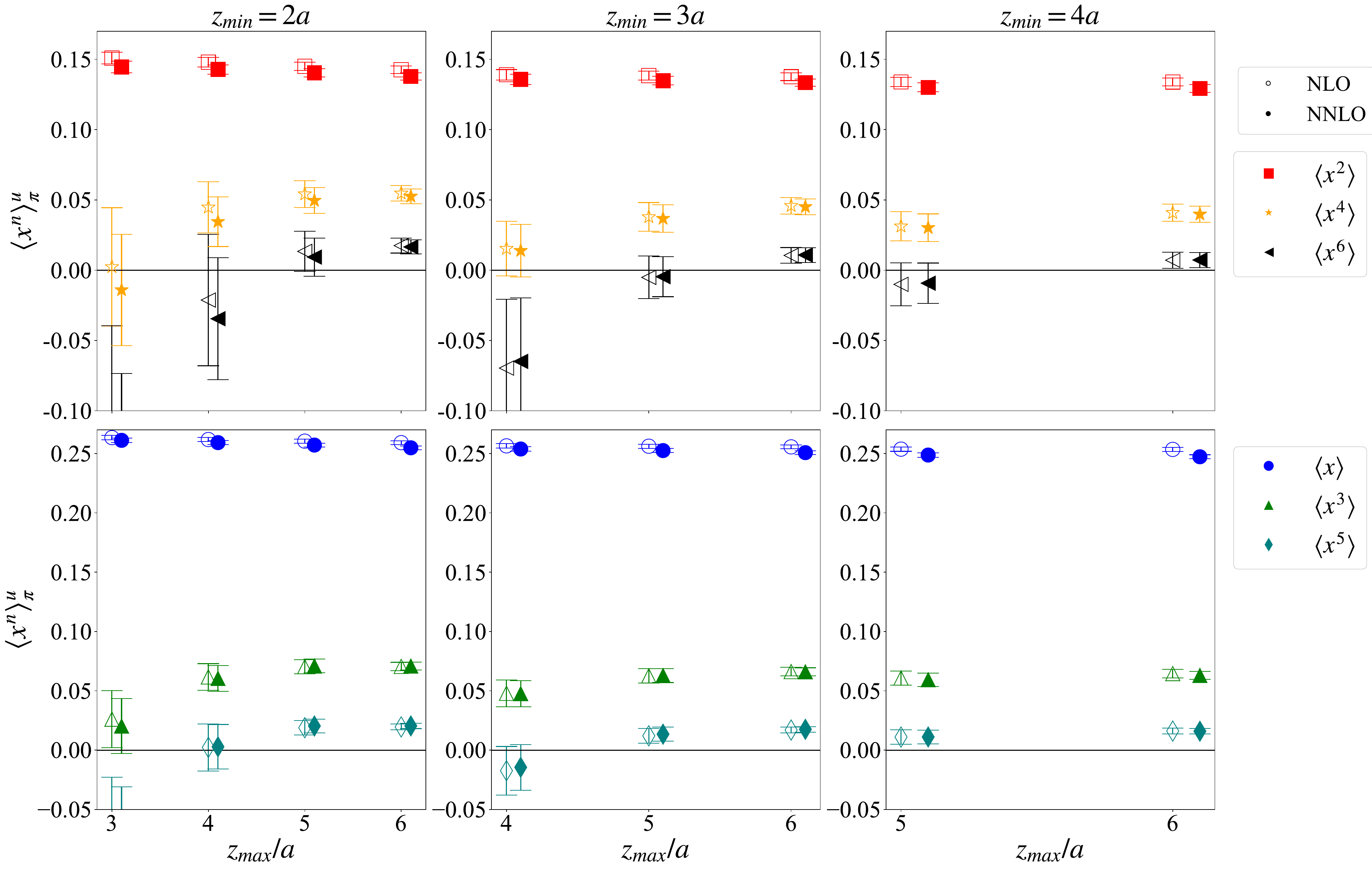}
    \vspace*{-0.2cm}
    \caption{Mellin moments $\langle x^n \rangle_\pi^u$ for $n=1,2,3,4,5,6$ at NLO (open symbols) and NNLO (filled symbols) accuracy using $z_{min} = 2a$ (left panel), $z_{min}{=}3a$ (middle panel), and $z_{min}{=}4a$ (right panel). In each case $z_{max}$ in the range $z_{min}+1$ to $6a$. }
    \label{fig:moments_pion}
\end{figure}

\begin{figure}[!h]
    \centering    \includegraphics[scale=0.237]{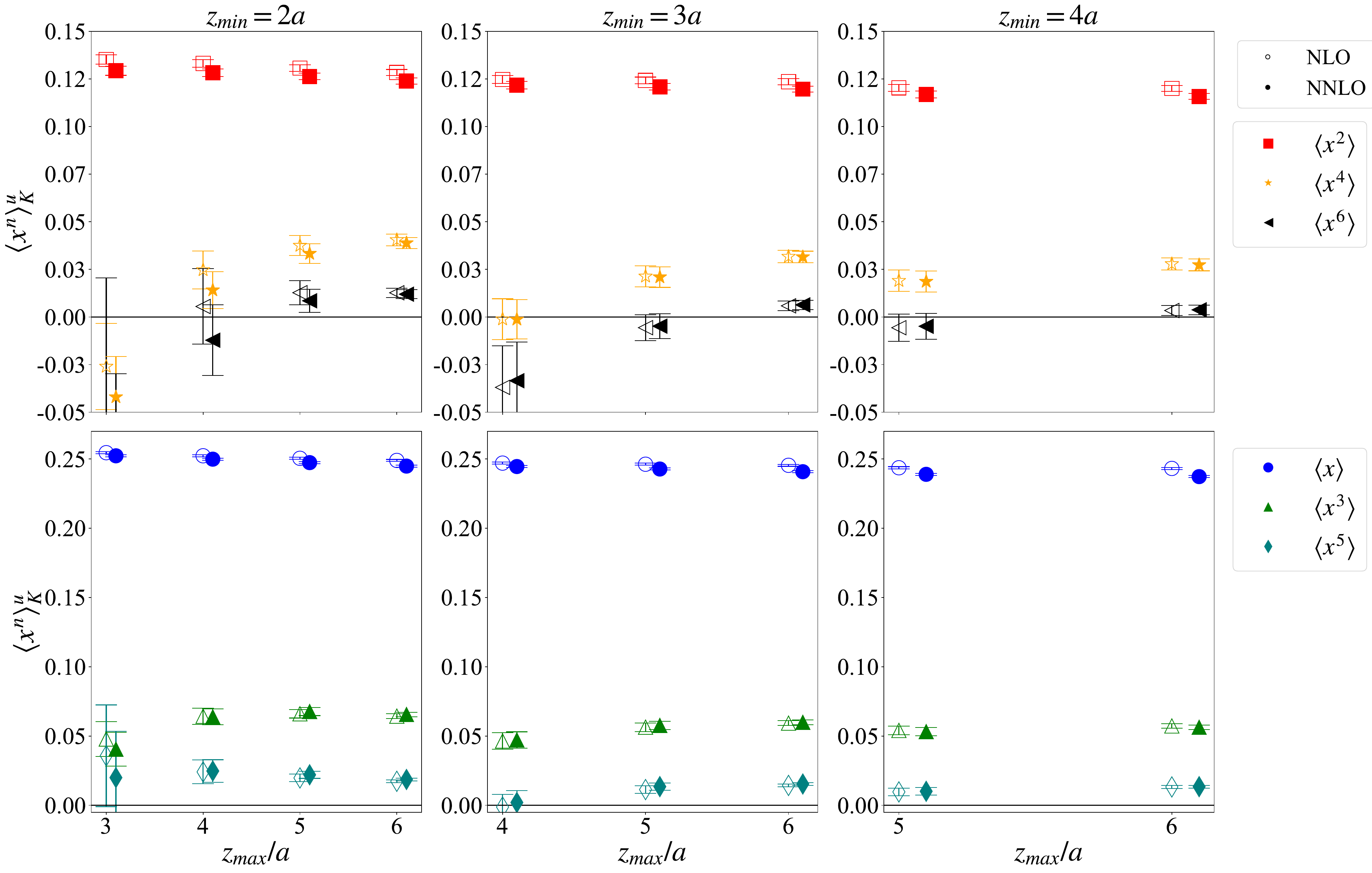}
    \vspace*{-0.2cm}
    \caption{Mellin moments $\langle x^n \rangle_K^u$ from the combined-fit analysis. The notation is the same as Fig.~\ref{fig:moments_pion}.}
    \label{fig:moments_kaon_u}
\end{figure}

\begin{figure}[!h]
    \centering
\includegraphics[scale=0.237]{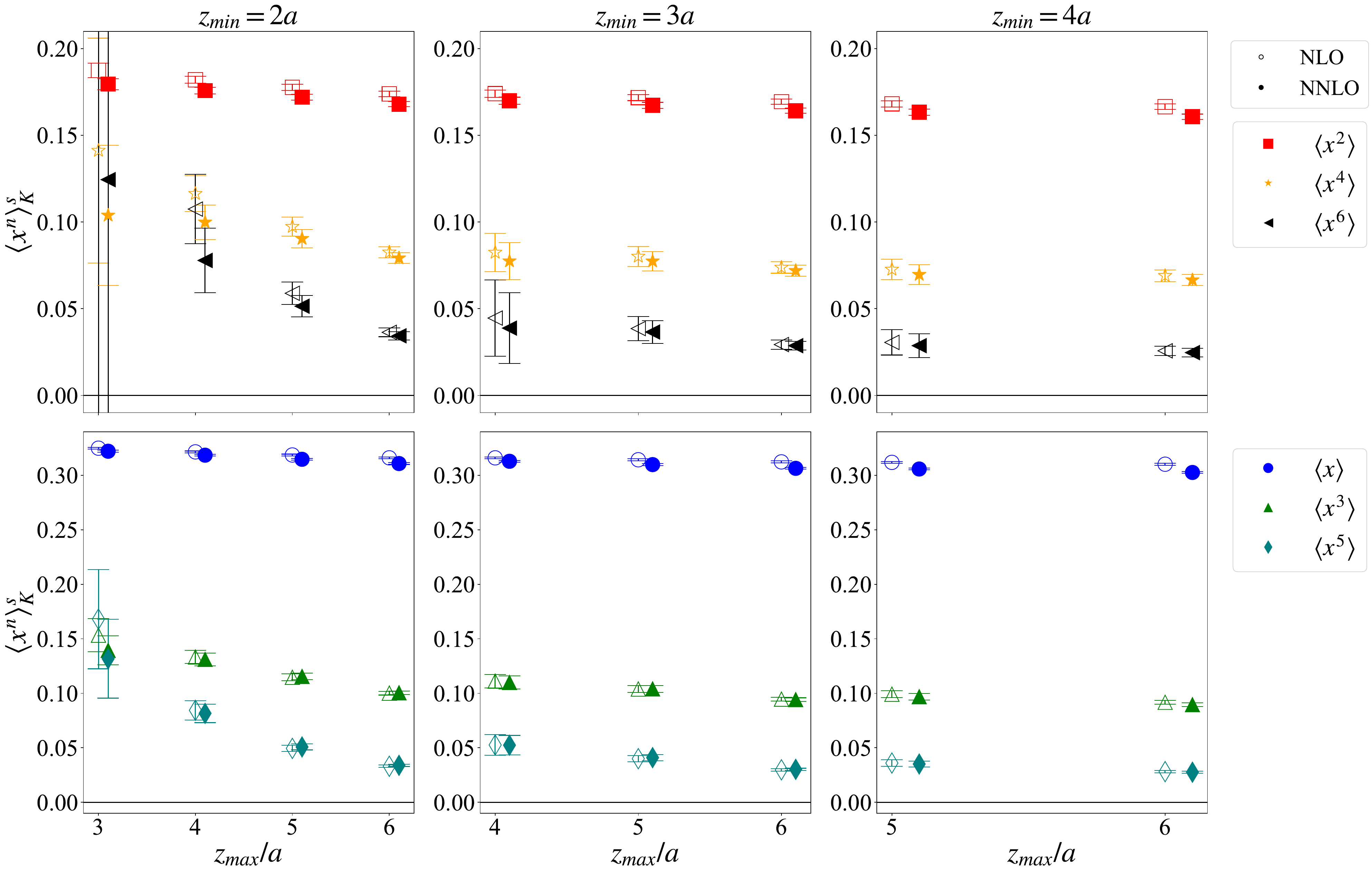}
        \vspace*{-0.2cm}
    \caption{Mellin moments $\langle x^n \rangle_K^s$ from the combined-fit analysis. The notation is the same as Fig.~\ref{fig:moments_pion}.}
    \label{fig:moments_kaon_s}
\end{figure}

One source of systematic uncertainty arises from the choice of the fit range in $z$. The short-distance expansion underlying Eq.~\eqref{eqn:moment_expansion} is expected to be valid within a window where discretization effects at small $z$ and higher-twist contributions at larger $z$ are both under control. To investigate this, we perform fits over multiple ranges with $z_{min}\in[2a,4a]$ and $z_{max}\in[3a,6a]$ with the constraint $z_{max}> z_{min}$, as $z_{max}=z_{min}$ corresponds to the analysis of Sec.~\ref{sub:z_fixed}. In this test, we use $n_{max} = 6$ and NLO accuracy. We exclude $z_{min}=1a$ from the analysis due to expected significant discretization effects at such short separations, and we include data up to $z=6a$. While the latter formally extends beyond the expected region of validity of the short-distance expansion, we nevertheless consider these separations in order to investigate their impact on the extracted moments within the statistical precision of the present calculation. 
Figs.~\ref{fig:moments_pion} - \ref{fig:moments_kaon_s} show the dependence of the extracted moments on the coordinate-space fit window at NLO accuracy. The three panels in each figure correspond to different choices of $z_{min}$, while the horizontal axis shows the variation with $z_{max}$. In the lowest moments, the results are relatively stable across the considered fit windows. In particular, changes in $z_{max}$ at fixed $z_{min}$ lead only to small variations, while the dependence on $z_{min}$ remains small but visible. The sensitivity to the fit window increases for higher moments. This is expected, since higher moments enter the OPE through higher powers of the Ioffe time and are therefore less strongly constrained by the available range of $(P_3,z)$ values. As a result, the moments with $n\geq 4$ exhibit larger statistical uncertainties and more noticeable fluctuations under changes of both $z_{min}$ and $z_{max}$. In Table~\ref{tab:NLO_moments} we report the Mellin moments from the NLO analysis along with a systematic uncertainty obtained from the variation of the $[z_{min},z_{max}]$ range. We note that these are not considered final results, as one can also obtain the moments at NNLO accuracy for the Wilson coefficients; final results can be found in Sec.~\ref{sub:final_results}.
\begin{table}[h!]
\begin{center}
\renewcommand{\arraystretch}{1.2}
\begin{tabular}{l|cccccc}
\hline
NLO & $\langle x \rangle$ & $\langle x^2 \rangle$ & $\langle x^3\rangle$ & $\langle x^4\rangle$ & $\langle x^5 \rangle$ & $\langle x^6 \rangle$  \\ \hline
$\pi^u$ & ~$0.255(2)(1)$~ & ~$0.136(3)(3)$~ & ~$0.064(4)(9)$~ & ~$0.040(7)(15)$~ & ~$0.016(3)(17)$~ & ~$0.006(7)(40)$~  \\\hline
$K^u$   & $0.245(1)(2)$ & $0.122(2)(2)$ & $0.057(2)(6)$ & $0.027(4)(16)$ & $0.013(1)(8)$ & $0.003(3)(21)$  \\\hline
$K^s$   & $0.313(1)(3)$ & $0.170(2)(4)$ & $0.096(2)(10)$ & $0.073(4)(7)$ & $0.030(1)(12)$ & $0.029(3)(9)$  \\\hline
\end{tabular}
\caption{Mellin moments at $\mu = 2$ GeV with NLO accuracy for $n_{max} = 6$ using a weighted average of the results with $z_{min}\in[3a,4a]$ and $z_{max}\in [4a,6a]$ constrained to $z_{max}-z_{min} \ge 1a$, with systematic errors corresponding to different fit ranges.}
\label{tab:NLO_moments}
\end{center}
\end{table}

We repeat the same analysis using NNLO Wilson coefficients, with the corresponding results shown with filled symbols in Figs.~\ref{fig:moments_pion} - \ref{fig:moments_kaon_s}. The qualitative behavior is similar to that of the NLO results. For all three distributions, the lowest moments remain comparatively stable under changes of the fit window, while the higher moments show increased sensitivity to the choice of $z_{min}$ and $z_{max}$. As in the NLO case, the dependence on $z_{max}$ at fixed $z_{min}$ is generally mild, whereas changing $z_{min}$ produces a more visible shift in the extracted moments. 
\newpage

\begin{figure}[!h]
    \centering
    \includegraphics[width=1\linewidth]{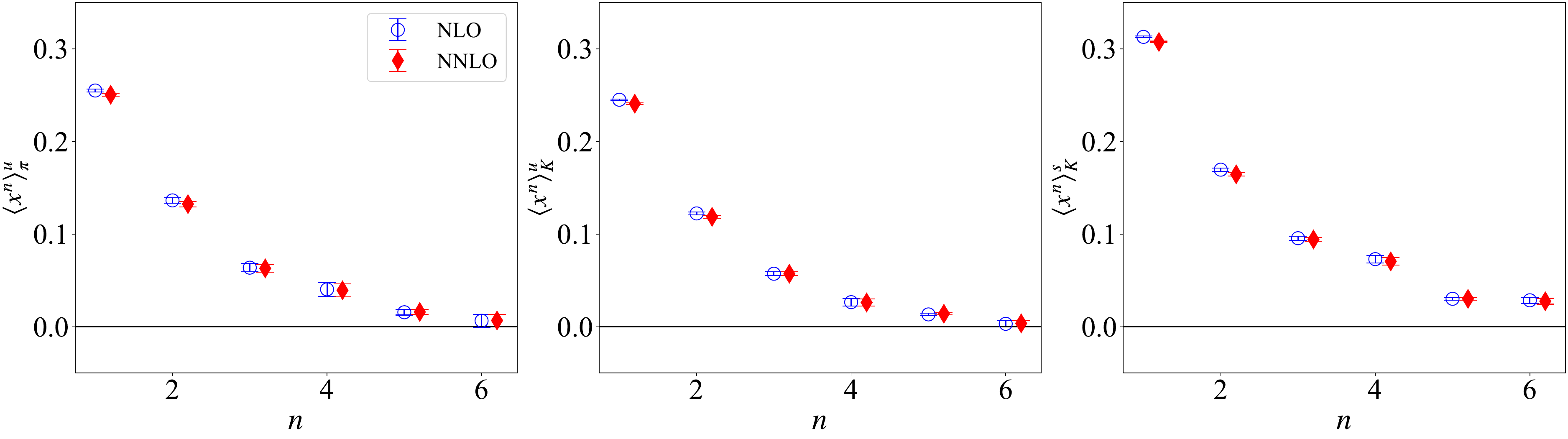}
    \caption{NLO (blue) vs NNLO (red) comparison for $\langle x^n\rangle^u_\pi$ (left), $\langle x^n\rangle^u_K$ (middle), and $\langle x^n\rangle^s_K$ (right) at for $n_{max} = 6$ using a weighted average of the results with $z_{min}\in[3a,4a]$ and $z_{max}\in [4a,6a]$ constrained to $z_{max}-z_{min} \ge 1a$. }
    \label{fig:NLO_NNLO}
\end{figure}

To summarize the dependence on the perturbative order, we construct weighted averages of the moments obtained from the combined $(P_3,z)$ fits over a selected set of fit windows. In particular, the averages include the fit windows with $z_{min}\in[3a,4a]$ and $z_{max}\in[z_{min}+a,6a]$, and use the statistical uncertainties of the individual extractions as weights. The averaging is performed separately at NLO and NNLO accuracy, and the resulting values are shown in Fig.~\ref{fig:NLO_NNLO} for both the pion and kaon. As can be seen, the inclusion of NNLO Wilson coefficients leads to a very small shift relative to NLO, with the effect slightly visible for the lower moments. Fig.~\ref{fig:NLO_NNLO} also exhibits the expected hierarchy among Mellin moments. The extracted values decrease with increasing moment order. This follows from the definition of the moments, since higher powers of $x$ suppress the small- and intermediate-$x$ regions and make the moments increasingly sensitive to the large-$x$ behavior of the PDF, which approaches zero. Consequently, the higher moments are more difficult to constrain and have larger relative uncertainties. 

\subsection{Renormalization-Group Evolution for Mellin moments}
\label{sub:evol}

The fixed-order analysis presented above evaluates the Wilson coefficients at the reference scale $\mu = 2$ GeV. This provides an extraction of the Mellin moments directly at the desired scale. However, the perturbative coefficients contain logarithms of the form $\ln(z^2\mu^2 e^{2\gamma_E}/4)$. When the scale $\mu$ is not close to the natural scale set by $z$, these logarithms may become numerically important. It is therefore useful to examine an alternative implementation in which the matching is performed at an initial scale $\mu_0$ associated with the short-distance separation, followed by perturbative evolution of the moments to $\mu=2$ GeV. The evolution relations are summarized in Appendix~\ref{appB}.

We define the initial scale as 
\begin{equation} 
\mu_0 = \frac{2\kappa}{z e^{\gamma_E}}\,, 
\end{equation} 
where the dimensionless parameter $\kappa$ controls the relation between the $z$ and the matching scale $\mu_0$. Note that the choice $\kappa=1$ removes the logarithm $\ln(z^2\mu_0^2e^{2\gamma_E}/4)$ from the Wilson coefficient. However, this choice does not necessarily minimize the residual scale dependence of the extracted moments, as one needs very fine lattices to access multiple values of $z$ within the perturbative regime. Thus, varying $\kappa$ provides a practical way to assess the sensitivity of the extraction to the perturbative scale entering the matching and evolution.

\begin{figure}[!h]
    \centering
    \includegraphics[scale=0.24]{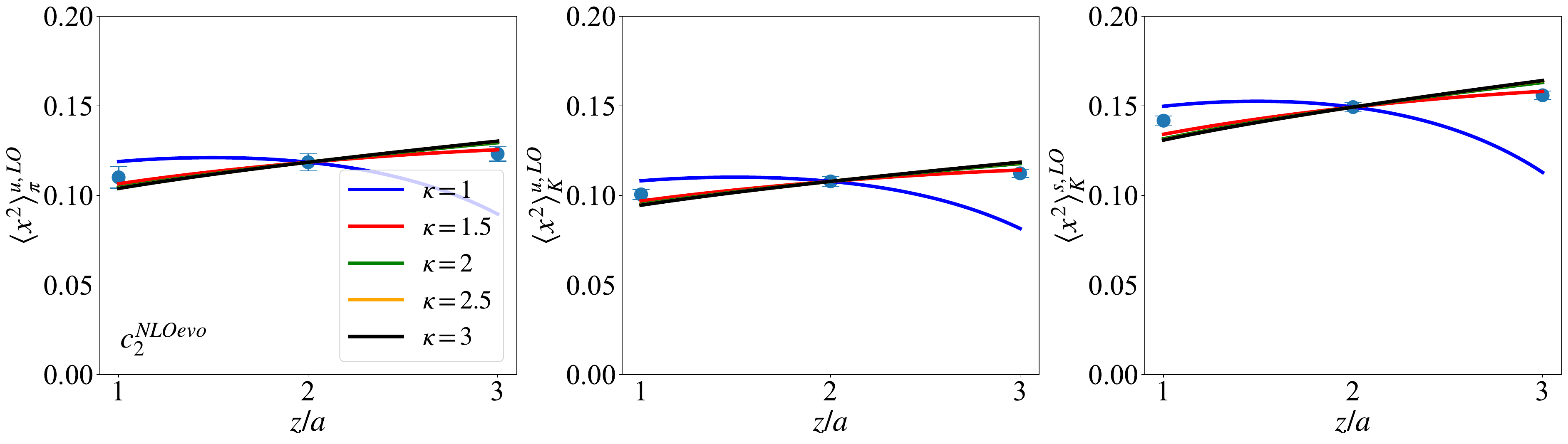}
    \includegraphics[scale=0.24]{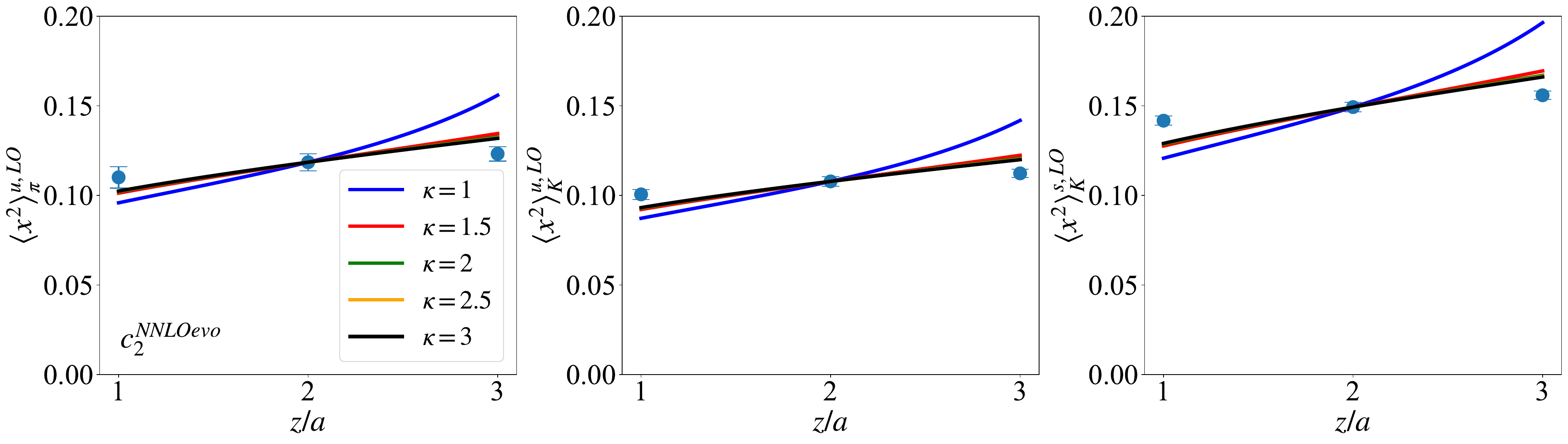}
    \vspace*{-0.35cm}
    \caption{Leading-order Mellin moment $\langle x^2 \rangle^{LO}$ for the pion (left), kaon up (middle), and kaon strange (right) components compared to the predictions from the NLO evolution (top) and NNLO evolution (bottom) Wilson coefficients for various $\kappa$. }
    \label{fig:LO_kappa_prediction}
\end{figure}

Following the strategy of Ref.~\cite{Gao:2022iex}, we construct the leading-order moments extracted from the lattice data, in which the Wilson coefficients are set to one, and compare their $z$ dependence with the behavior predicted by the Wilson coefficients upon evolution (RG-improved). This comparison is shown in Fig.~\ref{fig:LO_kappa_prediction}. We demonstrate the strategy on $\langle x^2\rangle_{M}^{f,{\rm LO}}$, which has a cleaner signal and better stability at small values of $z$. This is the region most relevant for the RG-improved analysis, since the initial scale $\mu_0\propto 1/z$ must remain sufficiently perturbative. The curves are normalized to the lattice result at $z=2a$, which serves as a reference separation; they are not fits to the data. Their purpose is to test whether the residual $z$ dependence of the fixed-$z$ extractions can be accounted for by perturbative evolution from an initial coordinate-space scale $\mu_0\propto \kappa/z$ to $\mu=2~{\rm GeV}$. The variation with $\kappa$ provides an estimate of the residual perturbative-scale dependence. We show data for $z\in[a,3a]$; however, the point at $z=a$ may be affected by sizable discretization effects, and our assessment focuses on $z=2a$ and $z=3a$.
The results in Fig.~\ref{fig:LO_kappa_prediction} show that the choice $\kappa=1$ does not describe the observed $z$ dependence well for either the NLO (NLOevo) or NNLO (NNLOevo) Wilson coefficients. Better agreement is obtained for larger values of $\kappa$. The dependence on $\kappa$ is slightly reduced from NLOevo to NNLOevo, indicating improved perturbative stability at NNLO. 

\begin{figure}[!h]
    \centering    
    \includegraphics[scale=0.23]{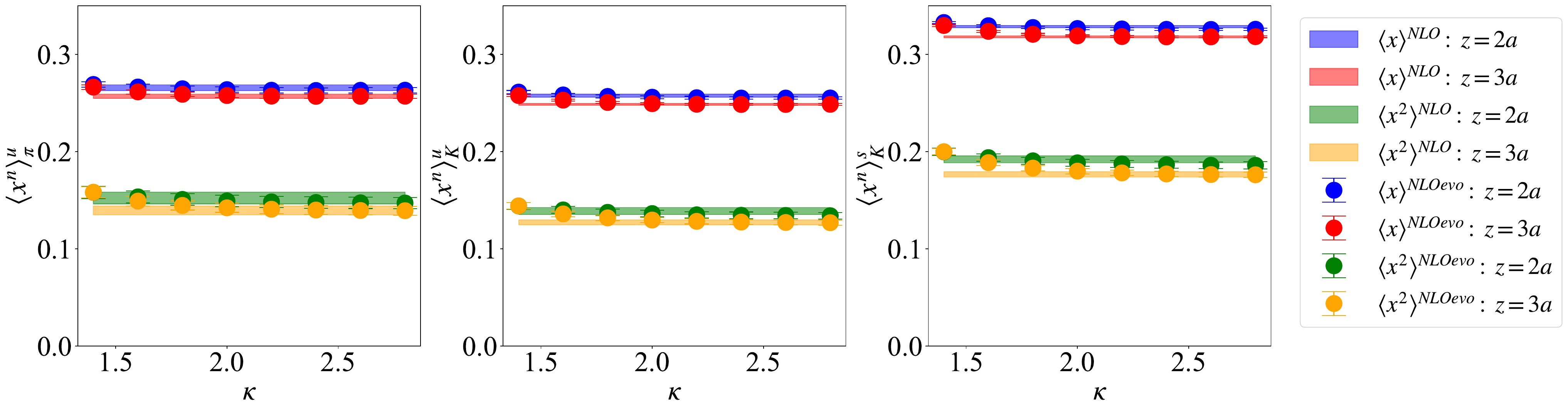}\\
    \includegraphics[scale=0.23]{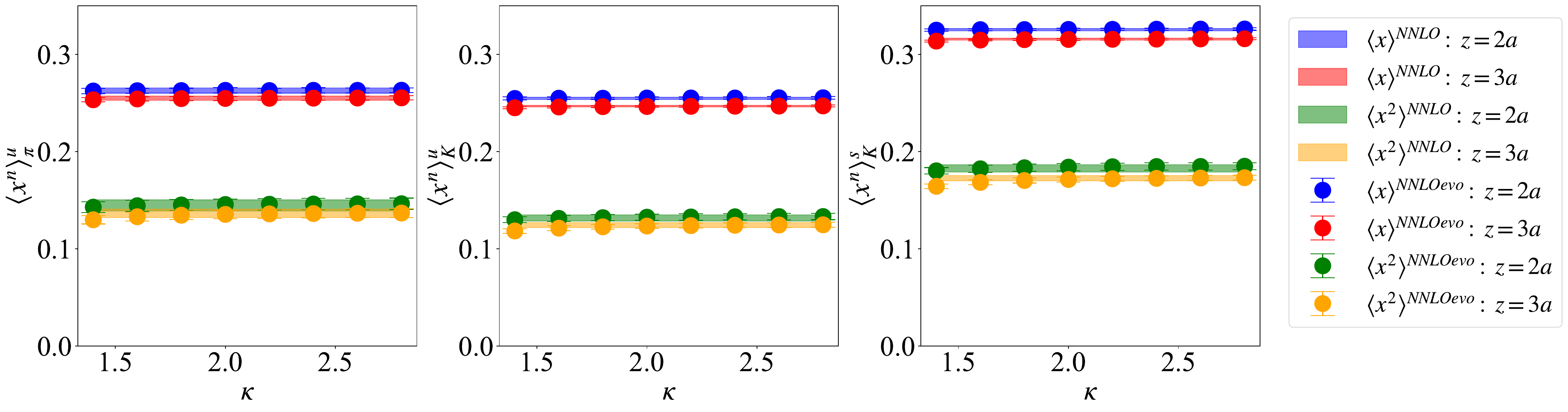}
        \vspace*{-0.35cm}
    \caption{Evolution of the moments $\langle x \rangle$ and $\langle x^2 \rangle$  at $z=2a$ (blue and green, respectively) and $z=3a$ (red and orange, respectively) at NLOevo (top) and NNLOevo (bottom) accuracy for the pion (left), kaon up-quark (middle), and kaon strange-quark (right). The fixed-scale determinations at $\mu=2~{\rm GeV}$ from Sec.~\ref{sub:fixed} are shown as bands for comparison. }
    \label{fig:NLOevo}
\end{figure}

Based on this behavior, we take $\kappa=2$ as the central value in the evolved analysis and vary $\kappa$ in the range $\kappa\in[\sqrt{2},2\sqrt{2}]$ to estimate the residual uncertainty related to the perturbative scale. In Fig.~\ref{fig:NLOevo}, we also show the results for the two lowest moments, $\langle x \rangle$ and $\langle x^2 \rangle$, obtained at $z=2a$ and $z=3a$ for both NLOevo and NNLOevo accuracy levels. The fixed-scale determinations at $\mu=2~{\rm GeV}$ presented in Sec.~\ref{sub:fixed}  (see Fig.~\ref{fig:fixed_z_moments_NLO_vs_NNLO}) are shown as bands for comparison. At NLOevo, the results show a dependence on $\kappa$ for $\kappa<2$ for both $z=2a$ and $z=3a$ and for all three flavor/meson combinations. The effect is more pronounced for $\langle x^2 \rangle$, which may indicate different $\kappa$ dependence in the real and imaginary parts of the double ratio. In contrast, the NNLOevo results show a much weaker dependence on $\kappa$, and, in addition, they are generally compatible with the fixed-order results within uncertainties. 

In Table.~\ref{tab:Evo_moments} we collect the results for NLOevo and NNLOevo combining the $z=2a$ and $z=3a$ cases. We report a systematic error arising from the $\kappa$ dependence, which, as shown, is larger in the NLOevo case. While the results discussed here support the stability of the RG evolution, extending this procedure to all moments considered in this work is challenging. This limitation arises because the RG evolution requires sufficiently small Wilson-line separations, where the initial scale remains perturbative. However, the higher moments are better constrained by fit windows that include larger values of $z$, as seen in Figs.~\ref{fig:moments_pion} - \ref{fig:moments_kaon_s}. 
\begin{table}[h!]
\begin{center}
\renewcommand{\arraystretch}{1.5}
\begin{tabular}{l|cccc}
\hline
Evolution & $\quad\langle x \rangle^{NLOevo}\quad$ & $\quad\langle x\rangle^{NNLOevo}\quad$ & $\quad\langle x^2 \rangle^{NLOevo}\quad$ & $\quad\langle x^2 \rangle^{NNLOevo}\quad$  \\ \hline
$\pi^u$ & ~0.263(2)(4)~ & ~0.261(2)(1)~ & ~0.150(4)(7)~ & ~0.145(4)(2)~ \\\hline
$K^u$   & 0.254(1)(3) & 0.252(1)(1) & 0.134(3)(7) & 0.130(2)(2) \\\hline
$K^s$   & 0.324(1)(4) & 0.322(1)(1) & 0.187(6)(9) & 0.180(3)(3) \\\hline
\end{tabular}
\caption{Mellin moments $\langle x \rangle$ and $\langle x^2\rangle$ evolved from $\mu_0 = \frac{4}{ze^{\gamma_E}}$ to $\mu=2$ GeV at NLOevo and NNLOevo accuracy with $n_{max}=6$ and a fit range over $z \in[2a,3a]$, with systematic errors corresponding to varying $\kappa \in[1.4,2,2.8]$.}
\label{tab:Evo_moments}
\end{center}
\end{table}

\subsection{Final Results}
\label{sub:final_results}

The analyses of the previous sections provide several complementary checks on the extraction of Mellin moments from the reduced Ioffe-time distributions. The fixed-$z$ fits illustrate the residual dependence on the Wilson-line separation, the combined $(P_3,z)$ fits improve the constraints by using multiple boosts and separations simultaneously, and the comparison between NLO and NNLO Wilson coefficients provides a measure of perturbative stability in the coefficients. In addition, the RG-improved analysis of Sec.~\ref{sub:evol} offers a useful consistency check for the lowest moments. Based on these studies, we adopt the moments obtained from the simultaneous $(P_3,z)$ fits at NNLO accuracy as our final results. We use the OPE truncation $n_{\rm max}=6$ and construct weighted averages over the fit windows with $z_{\rm min}\in[3a,4a]$ and $z_{\rm max}\in[4a,6a]$, subject to the condition $z_{\rm max}-z_{\rm min}\ge a$. The weights are determined from the statistical uncertainties of the individual fits. The use of this window avoids the shortest separation, $z=a$, where discretization effects are expected to be largest, while still retaining enough coordinate-space information to constrain the higher moments. 

The final values are quoted at $\mu=2~{\rm GeV}$ in the $\overline{\rm MS}$ scheme and are summarized in Table~\ref{tab:NNLO_moments}. The reported uncertainties are separated into three contributions. The first uncertainty is statistical. The second reflects dependence on the coordinate-space fit window and is obtained from the variation across the fit ranges that enter the weighted average. The third uncertainty estimates the residual perturbative-order dependence and is taken as the absolute difference between the NLO and NNLO central values. The latter is a conservative estimate, as the difference between NNLO and NNNLO is expected to be smaller than the difference reported. This separation allows us to identify the dominant source of uncertainty for each moment. For the lowest moments, the statistical uncertainties are small, and the systematic effects from the fit window and perturbative order become important. For higher moments, the uncertainty is dominated by the increasing sensitivity to the fit range.

\bigskip
\begin{table}[h!]
\begin{center}
\renewcommand{\arraystretch}{1.5}
\begin{tabular}{l!{\vrule width 1.1pt}cccccc}
\Xhline{1.1pt}
NNLO & $\qquad\quad\langle x \rangle\qquad\quad$ & $\qquad\quad\langle x^2 \rangle\qquad\quad$ & $\qquad\quad\langle x^3\rangle\qquad\quad$ & $\qquad\quad\langle x^4\rangle\qquad\quad$ & $\qquad\quad\langle x^5 \rangle\qquad\quad$ & $\qquad\quad\langle x^6 \rangle\qquad\quad$  \\ \Xhline{1.1pt}
$\pi^u$ & $0.251(2)(3)(4)$ & $0.132(3)(3)(4)$ & $0.063(4)(09)(1)$ & $0.039(7)(16)(1)$ & $0.016(3)(16)(0)$ & $0.007(7)(38)(1)$ \\\hline
$K^u$   & $0.241(1)(4)(4)$ & $0.119(2)(3)(3)$ & $0.057(2)(06)(0)$ & $0.026(4)(16)(1)$ & $0.014(1)(07)(1)$ & $0.004(3)(20)(1)$ \\\hline
$K^s$   & $0.308(1)(5)(5)$ & $0.165(2)(5)(5)$ & $0.094(2)(10)(2)$ & $0.071(4)(05)(2)$ & $0.030(1)(12)(0)$ & $0.028(3)(07)(1)$ \\\Xhline{1.1pt}\end{tabular}
\caption{Final results for the Mellin moments $\langle x^n \rangle$ at NNLO accuracy for $n_{max} = 6$ for the pion, kaon up, and kaon strange components using a weighted average of the results with $z_{min}\in[3a,4a]$ and $z_{max}\in [4a,6a]$ constrained to $z_{max}-z_{min} \ge 1a$. Results are presented at a scale of $\mu=2$ GeV. The error reported in the first parenthesis is statistical, the second is the systematic uncertainty due to different fit ranges, and the third is the difference between NLO and NNLO.}
\label{tab:NNLO_moments}
\end{center}
\end{table}

The final results show a clear hierarchy among the three distributions. The pion and kaon up-quark moments are close to each other for all moment orders considered, with a tendency for $\langle x^n \rangle_K^u$ to be lower, while the kaon strange-quark moments are systematically larger. Overall, the final results show that nonlocal matrix elements, analyzed through the short-distance expansion, provide access to multiple Mellin moments of the pion and kaon PDFs within a unified framework. The lowest moments are well constrained, while the higher moments remain more sensitive to the choice of fit window and should be interpreted with greater caution. Nevertheless, the extraction of moments up to $\langle x^6\rangle$ illustrates the potential of the nonlocal-operator approach for accessing higher moments that are difficult to obtain from local operators.

\subsection{PDF Reconstruction}
\label{sec:reconstruction}

The Mellin moments provide integral constraints on the underlying parton distributions. Although a finite number of moments does not uniquely determine a PDF, it can be used to reconstruct its features within a functional form. Here, we examine the extent to which the moments extracted from the nonlocal matrix elements constrain the pion and kaon distributions.
With the convention adopted in Sec.~\ref{sec:theory}, the even moments probe the valence combination $q-\bar q$, whereas the odd moments probe $q+\bar q=q_v+2q_s$. The present calculation includes connected diagrams only. We therefore combine all moments into a single reconstruction that yields the valence combination, assuming the omitted disconnected contribution is small.
We parametrize the valence distribution of quark flavor $f$ in meson $M$ using the normalized two-parameter form
\begin{equation}
q_M^f(x)
=
\frac{x^\alpha(1-x)^\beta}
     {B(\alpha+1,\beta+1)},
\label{eq:pdf_ansatz}
\end{equation}
where $B$ is the Euler beta function. The normalization is chosen such that
\begin{equation}
\int_0^1 dx\, q_M^f(x)=1.
\end{equation}
The Mellin moments generated by this parametrization are
\begin{equation}
\langle x^n\rangle
=
\int_0^1 dx\,x^n q_M^f(x)
=
\frac{B(\alpha+n+1,\beta+1)}
     {B(\alpha+1,\beta+1)}.
\label{eq:beta_moments}
\end{equation}

\begin{figure}[!h]
    \centering    \includegraphics[scale=0.22]{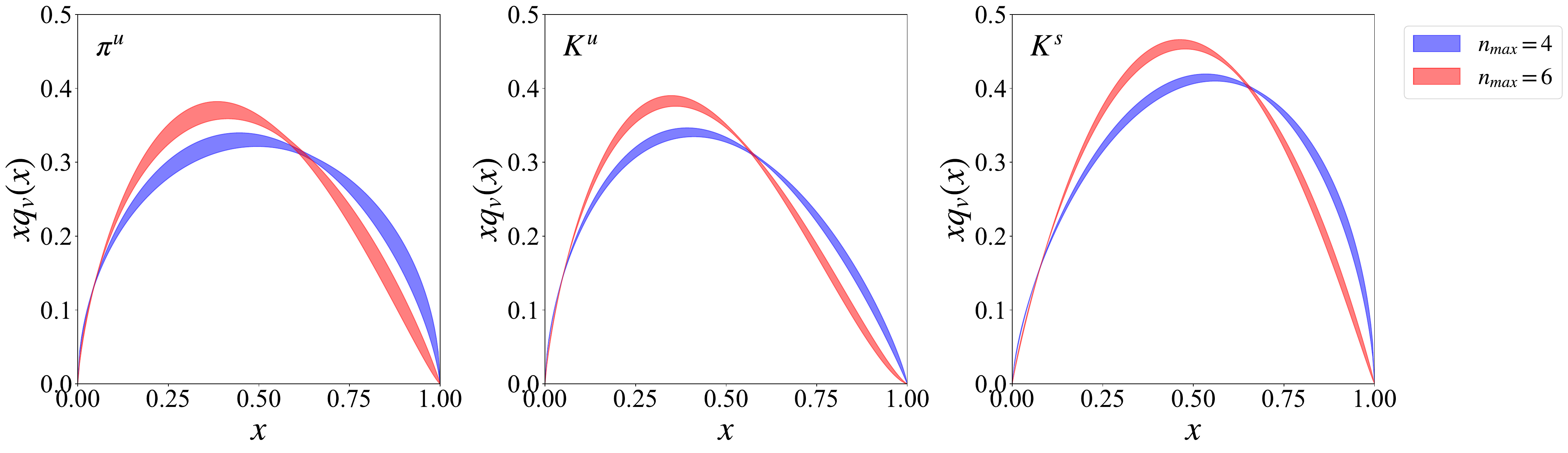}
    \vspace*{-0.3cm}
    \caption{Reconstructed distributions $xq(x)$ at $\mu{=}2$ GeV using moments up to $n_{max}=4$ (blue band) and $n_{max}=6$ (red band).}
    \label{fig:xqv_reconstruction}
\end{figure}

To assess the sensitivity of the reconstruction to the number of included moments, we fit the moments up to $n_{max}=4$ and 6. The corresponding distributions are shown in Fig.~\ref{fig:xqv_reconstruction}, and the fitted $\alpha$ and $\beta$ are given in Table~\ref{tab:pdf_fit_parameters}. 
\begin{table}[h!]
\begin{center}
\renewcommand{\arraystretch}{1.2}
\begin{tabular}{l|cc}
\hline
$n_{max}$ & 4 & 6   \\ \hline
$\pi^u$ & $\{-0.46(4),0.62(10)\}$ & $\{-0.31(5),1.05(13)\}$  \\\hline
$K^u$   & $\{-0.39(2),0.90(7)\}$ &  $\{-0.24(3),1.39(9)\}$  \\\hline
$K^s$   & $\{-0.29(2),0.59(5)\}$ & $\{-0.10(3),1.02(7)\}$  \\\hline\end{tabular}
\caption{Fitted parameters $\{\alpha,\beta\}$ of the valence distribution for different maximum moments ($n_{max}$) entering the fit. Values in parentheses are the statistical error.}
\label{tab:pdf_fit_parameters}
\end{center}
\end{table}

\begin{figure}[!h]
    \centering    
    \includegraphics[scale=0.26]{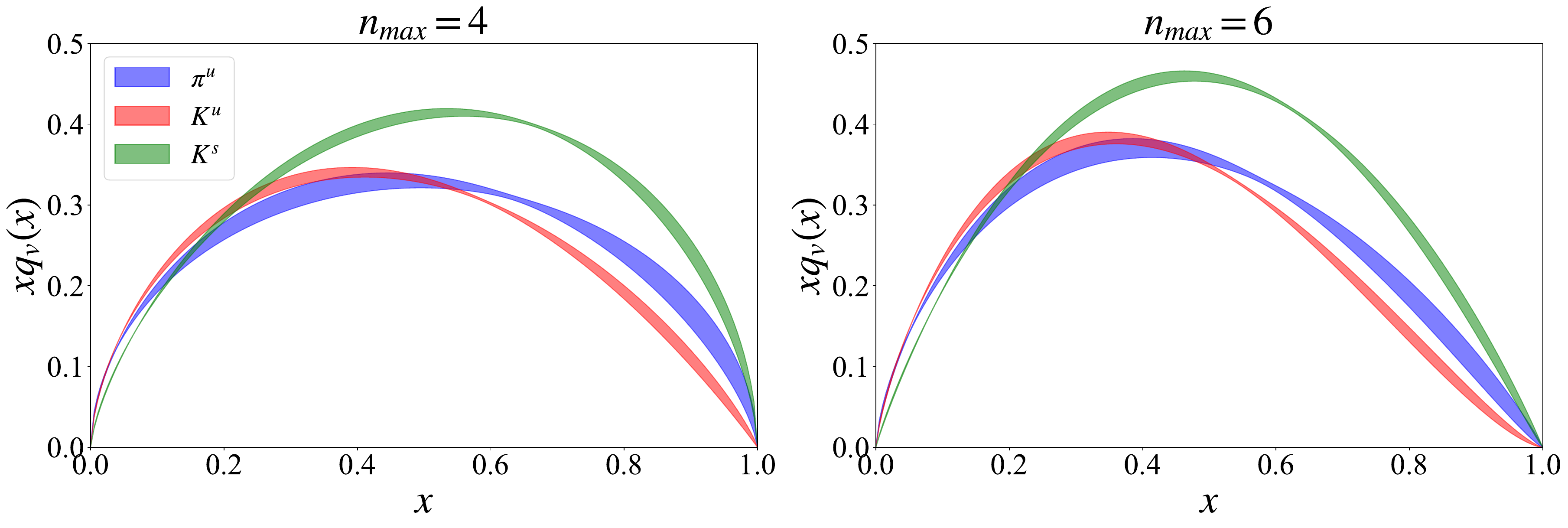}
    \vspace*{-0.25cm}
    \caption{Comparison of the pion (blue), kaon up-quark (red), and kaon strange-quark (green) distributions with the maximum moment entering the fit being $\langle x^4 \rangle$ (left) and $\langle x^6\rangle$ (right) at $\mu = 2$ GeV.}
    \label{fig:xqv_particle_comparison}
\end{figure}

Fig.~\ref{fig:xqv_particle_comparison} shows a comparison of the reconstructed valence distribution for the two particles. For all three cases, including moments through $\langle x^6\rangle$ modifies the shape relative to the reconstruction based on moments through $\langle x^4\rangle$. The peak moves toward smaller $x$, while the distribution becomes more suppressed at large $x$. This reflects the additional constraints imposed by the fifth and sixth moments, although these moments also carry substantially larger uncertainties and are more sensitive to the coordinate-space fit window. The effect of increasing $n_{\max}$ is therefore not negligible, demonstrating that the detailed reconstructed shape remains sensitive to the highest moments included in the fit. For a fixed $n_{\max}$ the pion and kaon up-quark distributions have similar shapes, with the kaon up-quark distribution shifted slightly toward smaller $x$. In contrast, the kaon strange-quark distribution peaks at a larger value of $x$ and remains enhanced toward the large-$x$ region. This behavior is consistent with the hierarchy observed in the Mellin moments, for which the strange-quark moments of the kaon are systematically larger than the corresponding pion and kaon up-quark moments. 

\section{Comparisons with other studies}
\label{sec:comparison}

It is useful to compare the moments extracted in this work with previous determinations obtained on the same lattice ensemble using local operators. Such a comparison provides a nontrivial cross-check, since the two approaches use different matrix elements and analysis strategies. In the local-operator method, each Mellin moment is obtained from the matrix element of a corresponding local operator (which contains covariant derivatives), whereas in the present work, the moments are extracted from the short-distance behavior of nonlocal matrix elements via the OPE. Using the same ensemble removes several sources of ambiguity in the comparison, including the lattice spacing, pion mass, and finite-volume setup. Remaining differences, therefore, primarily reflect the operator construction, renormalization, and matching procedures, excited-state treatment, and the fitting strategy used to isolate the moments.

\begin{table}[h!]
\begin{center}
\renewcommand{\arraystretch}{1.45}
\begin{tabular}{l|cc|cc|cc}
\hline
& \multicolumn{2}{c|}{$\langle x\rangle$}
& \multicolumn{2}{c|}{$\langle x^2\rangle$}
& \multicolumn{2}{c}{$\langle x^3\rangle$} \\
\cline{2-7}
& $\quad$Nonlocal\,\, & \,\, Local~\cite{Alexandrou:2020gxs}$\quad$
& $\quad$Nonlocal\,\, & \,\, Local~\cite{PhysRevD.104.054504}$\quad$
& $\quad$Nonlocal\,\, & \,\, Local~\cite{PhysRevD.104.054504}$\quad$ \\
\hline
$\pi^u$
& $0.251(5)$ & $0.273(9)$ & $0.132(6)$ & $0.110(6)$ & $0.063(10)$ & $0.026(17)$ \\ \hline
$K^u$ & $0.241(6)$ & $0.262(3)$ & $0.119(5)$ & $0.101(2)$ & $0.057(6)$ & $0.043(7)$ \\ \hline 
$K^s$ & $0.308(7)$ & $0.332(3)$ & $0.165(7)$ & $0.145(2)$ & $0.094(10)$ & $0.079(6)$ \\
\hline
\end{tabular}
\caption{Comparison of the Mellin moments $\langle x\rangle$ - $\langle x^3\rangle$ obtained from nonlocal operators in this work (Table~\ref{tab:NNLO_moments}) and from local-operator calculations using the same ensemble as this work~\cite{Alexandrou:2020gxs,PhysRevD.104.054504}. All results are quoted at $\mu=2$ GeV. For the nonlocal-operator results, the statistical, fit-range systematic, and perturbative-order uncertainties have been combined in quadrature.}
\label{tab:local_nonlocal}
\end{center}
\end{table}
Table~\ref{tab:local_nonlocal} summarizes the comparison for the first three moments of the pion up-quark distribution, the kaon up-quark distribution, and the kaon strange-quark distribution. The local-operator results for $\langle x\rangle$ are taken from Ref.~\cite{Alexandrou:2020gxs}, while those for $\langle x^2\rangle$ and $\langle x^3\rangle$ are taken from Ref.~\cite{PhysRevD.104.054504}. All results are quoted at $\mu=2$ GeV. It is noteworthy that for the local-operator approach, accessing $\langle x^3\rangle$ is already extremely challenging. One complication is that it requires higher-derivative operators whose matrix elements are more sensitive to many systematic effects, and the renormalization is a nontrivial process. In fact, to avoid power-divergent mixing requires momentum-boosted states with two nonzero spatial components for $\langle x^2\rangle$ and three for $\langle x^3\rangle$. Thus, the difference between the local and nonlocal calculations for the pion $\langle x^3\rangle$ is not surprising. Overall, we find qualitative consistency between the local- and nonlocal-operator determinations. For the first moment, the nonlocal-operator results have central values that are lower than the corresponding local-operator determinations for all three cases. For the second moment, the central values obtained in this work are larger than those from the local-operator calculation, with the largest shifts appearing for the pion and kaon up-quark moments. When uncertainties are taken into account, the remaining non-overlap is typically below the $10\%$ level, except for the pion $\langle x^3\rangle$, where the extraction is intrinsically more difficult. We also note that the extraction from the local operators contains only statistical errors.

Given the larger number of lattice-QCD studies employing nonlocal operators for the pion than for the kaon, we compare our pion moments with other determinations obtained at the same or a nearby renormalization scale, including scales up to 3.2 GeV. The comparison is summarized in Table~\ref{tab:pion_moments_lit_comparison}, where the first five columns list the main parameters of each calculation, and the last four columns show the corresponding values of $\langle x\rangle_\pi^u$ through $\langle x^4\rangle_\pi^u$ graphically. The vertical bands indicate the total uncertainty of the present work, where the statistical, fit-window systematic, and perturbative-order uncertainties have been combined in quadrature.
The results of Ref.~\cite{Gao:2025inf} are extracted directly from the short-distance behavior of ratios of nonlocal matrix elements using the appropriate Wilson coefficients (``OPE''). By contrast, Refs.~\cite{Lin:2020ssv,Gao:2020ito} first reconstruct the $x$-dependent pion PDF and then determine its Mellin moments from the integral definition (``PDF fit''). Ref.~\cite{Joo:2019bzr} employs both strategies on ensembles with $m_\pi\simeq415$ MeV and $a=0.127$ fm; the row marked with a star corresponds to the PDF-fit moments obtained using two ensembles with different volumes. More specifically, the LaMET-based analysis of Ref.~\cite{Lin:2020ssv} uses three pion masses and two lattice spacings and quotes moments derived from the reconstructed pion PDF, while Ref.~\cite{Gao:2020ito} provides a PDF-fit determination from the $a=0.06$ fm ensemble at $m_\pi\simeq300$ MeV. For Ref.~\cite{Gao:2025inf}, we quote the even moments at zero momentum transfer from the fixed-order $\mu=2$ GeV bands and include the difference from the NNL0-resummed result as an additional systematic uncertainty. The OPE and PDF-fit rows should therefore be viewed as complementary determinations rather than as strictly equivalent extractions.

Overall, the comparison shows broad consistency among the different nonlocal-operator approaches. The lowest moment, $\langle x\rangle_u^\pi$, is generally found in the range $0.19$--$0.28$, with the spread reflecting differences in pion mass, lattice spacing, reconstruction strategy, perturbative matching, and analysis systematics. Our value lies well within the range of previous determinations. For $\langle x^2\rangle_u^\pi$, the various determinations have a wider spread. We note, however, that systematic uncertainties are not reported in all calculations and, when included, may account for different sources of uncertainty. Finally, $\langle x^3\rangle_u^\pi$ and $\langle x^4\rangle_u^\pi$ are compatible between the different calculations within the reported uncertainties, which are, in most cases, higher than the lowest moments.
\begin{table}[h!] 
\centering 
\renewcommand{\arraystretch}{1.75} 
{\small{
\begin{minipage}[c]{0.26\textwidth}
\centering 
\begin{tabular}{llccc}
\, & \hspace*{-0.25cm} Method & $m_\pi$\,(MeV) & $a$\,(fm) & $\mu$\,(GeV) \\ \hline
This work & OPE & 260 & 0.0934 & 2 \\ 
Ref.~\cite{Joo:2019bzr} & OPE & 415 & 0.127 & 2 \\ 
Ref.~\cite{Gao:2025inf} & OPE & 300 & 0.04 & 2 \\[.65ex] 
\cdashline{1-5}
Ref.~\cite{Joo:2019bzr} & PDF fit & 415 & 0.127\,$^\star$ & 2 \\ 
Ref.~\cite{Lin:2020ssv} & PDF fit & 220--320 & 0.06--0.12 & 2.4 \\ 
Ref.~\cite{Gao:2020ito} & PDF fit & 300 & 0.06 & 3.2 \\ 
\hline 
\end{tabular} 
\end{minipage}
}}
\hfill 
\begin{minipage}[c]{0.58\textwidth} 
\centering
\begin{tabular}{llll}
\,\\[-2ex]
\hspace*{-.5cm} $\langle x\rangle^u_\pi$ \hspace*{.75cm} 
&\hspace*{.95cm} $\langle x^2\rangle^u_\pi$  \hspace*{.75cm} 
&\hspace*{.95cm} $\langle x^3\rangle^u_\pi$  \hspace*{.75cm} 
&\hspace*{.75cm} $\langle x^4\rangle^u_\pi$  \\[-0.05cm]
\end{tabular}
\hspace*{-0.32cm}
\includegraphics[scale=0.165]{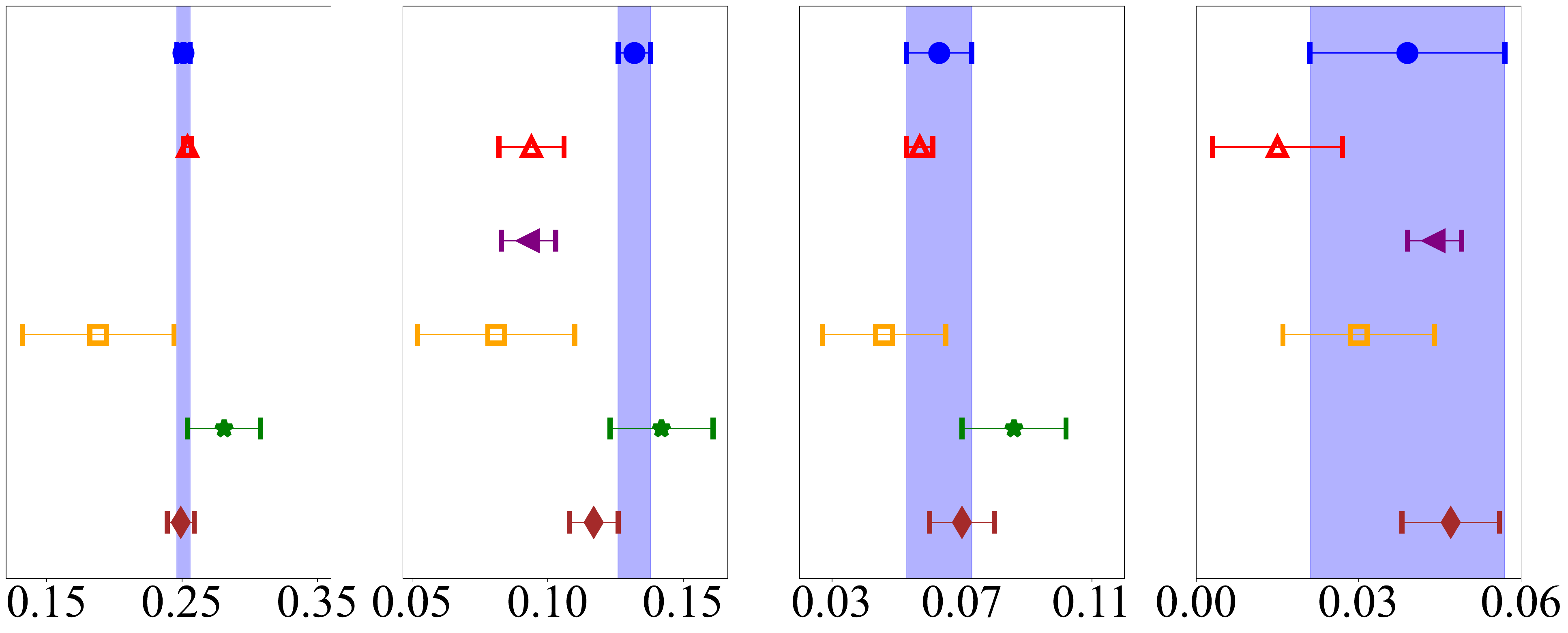} 
\end{minipage} 
\caption{Comparison of pion up-quark Mellin moments, $\langle x^n\rangle^u_\pi$, from calculations using a nonlocal-operator. Filled symbols indicate that statistical and systematic uncertainties are combined in quadrature, while open symbols only have statistical uncertainties available. $^\star$ The calculation combines two ensembles at different volumes. } 
\label{tab:pion_moments_lit_comparison} 
\end{table}

\begin{figure}[!h]
    \centering    \includegraphics[width=0.62\linewidth]{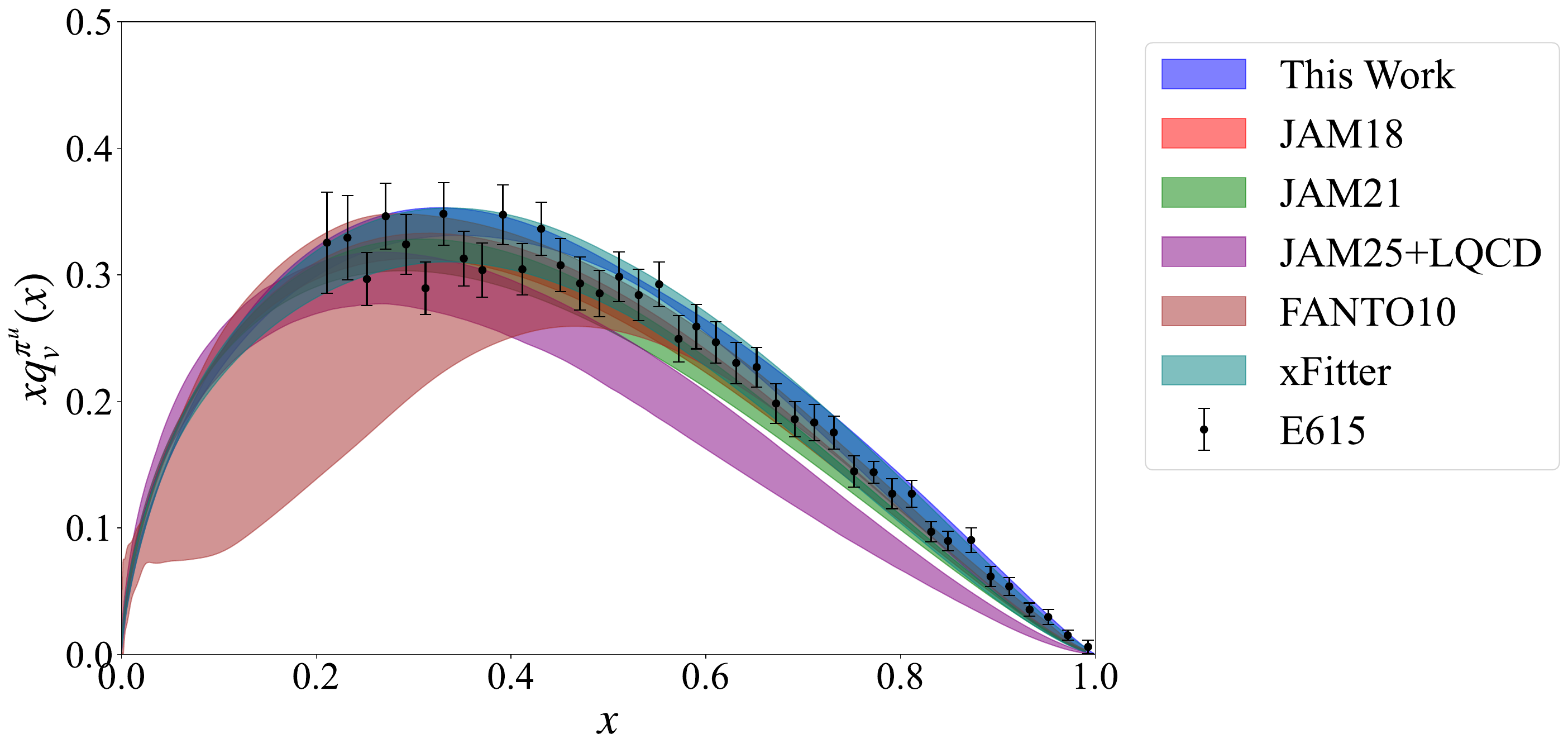}
    \vspace*{-0.35cm}
    \caption{Pion $xq_v$ from this work (blue band) compared to the phenomenological estimates of JAM18~\cite{Barry:2018ort} (red band), xFitter~\cite{Novikov:2020snp} (turquoise band),  JAM21~\cite{Barry:2021osv} (green band), FANTO10~\cite{Kotz:2025lio} (brown band), the JAM25 with lattice-QCD constraints~\cite{Barry:2025wjx} (purple band), and the Fermilab E615 experiment~\cite{Conway:1989fs} (black points). Results are shown at a scale of $\mu = 5.2$ GeV.}
    \label{fig:5.2GeV_pi_comparison}
\end{figure}

\begin{figure}[!h]
    \centering    \includegraphics[width=0.62\linewidth]{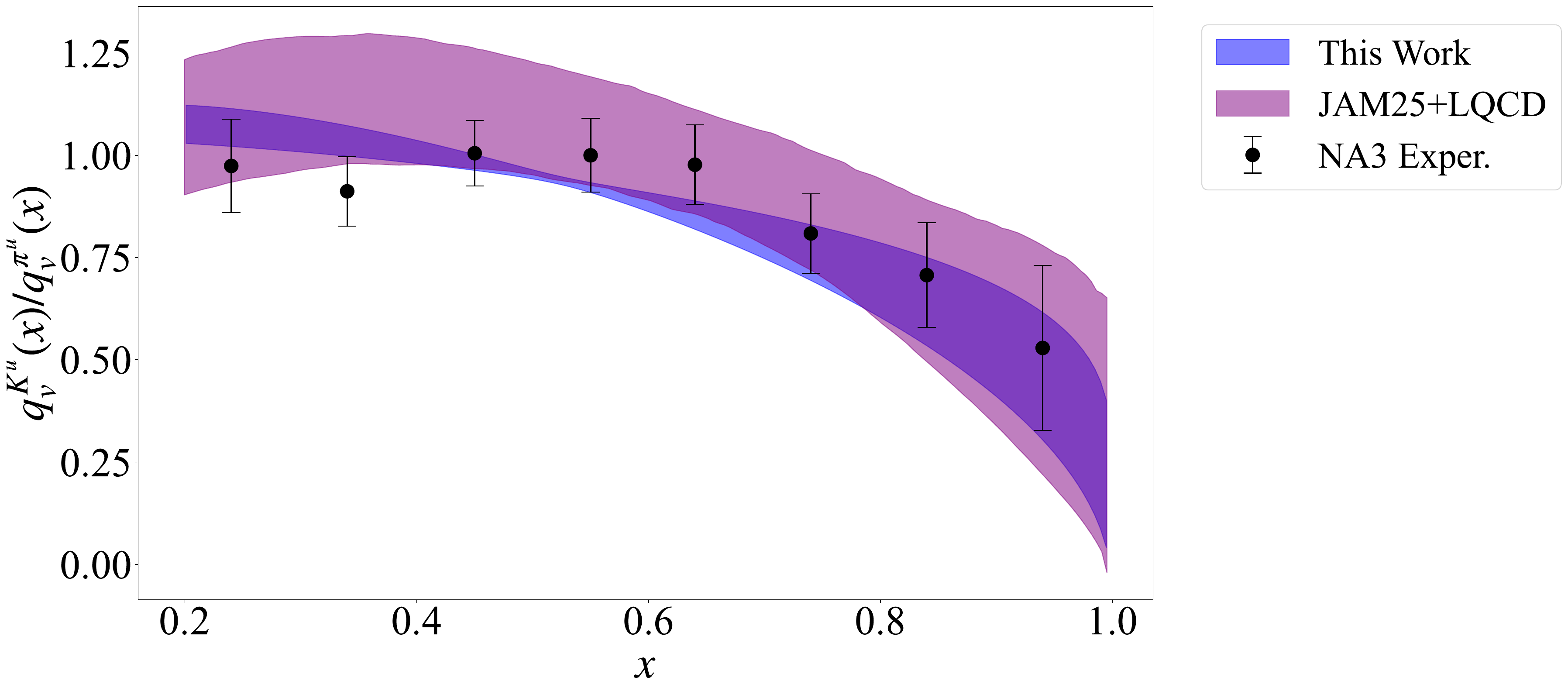}
    \vspace*{-0.35cm}    
    \caption{Ratio of kaon up-quark to pion strange reconstructed PDFs (blue band) and CERN-NA3 experimental data~\cite{BADIER1980354} and JAM25+LQCD~\cite{Barry:2025wjx} at $\mu = 5.2$ GeV.}
    \label{fig:kaon_u_pion_ratio_exp_comparison}
\end{figure}

Having obtained the reconstructed pion and kaon PDFs, it is interesting to compare them with phenomenological determinations. For this comparison, all results are evolved to $\mu=5.2~\mathrm{GeV}$ at the $\overline{\mathrm{MS}}$ scheme, which is customary. Our results for the pion valence PDF are shown in Fig.~\ref{fig:5.2GeV_pi_comparison}, with the phenomenological estimates of JAM18~\cite{Barry:2018ort}, xFitter~\cite{Novikov:2020snp},  JAM21~\cite{Barry:2021osv}, FANTO10~\cite{Kotz:2025lio}, as well as the recent JAM 2025 analysis incorporating lattice-QCD constraints~\cite{Barry:2025wjx}. We also include data from the Fermilab E615 experiment~\cite{Conway:1989fs}. We find excellent compatibility with the phenomenological estimates and the E615 data, except for the JAM25+LQCD band, which is lower than the other results in the $x$ region after $xq_v$ peaks. We note that the JAM25 results incorporate next-to-leading-logarithmic (NLL) threshold resummation and extract the pion and kaon distributions simultaneously. For the kaon valence distributions, we compare our results with the recent data of JAM25+LQCD. The strange-quark contribution is fully compatible between the two results, and in the up-quark case, there is a very good agreement, with a minor tension in the intermediate $x$ region. Nevertheless, this agreement is very promising, as the lattice data are at a pion mass of 260 MeV, and there are systematic uncertainties to be quantified. Fig.~\ref{fig:kaon_u_pion_ratio_exp_comparison} shows the ratio of the up-quark distributions in the kaon and pion at $\mu=5.2~\mathrm{GeV}$, which has also been measured in the NA3 experiment. The JAM25+LQCD is also shown for comparison.  As can be seen, all results are in excellent agreement.
\begin{figure}[!h]
    \centering    \includegraphics[width=0.8\linewidth]{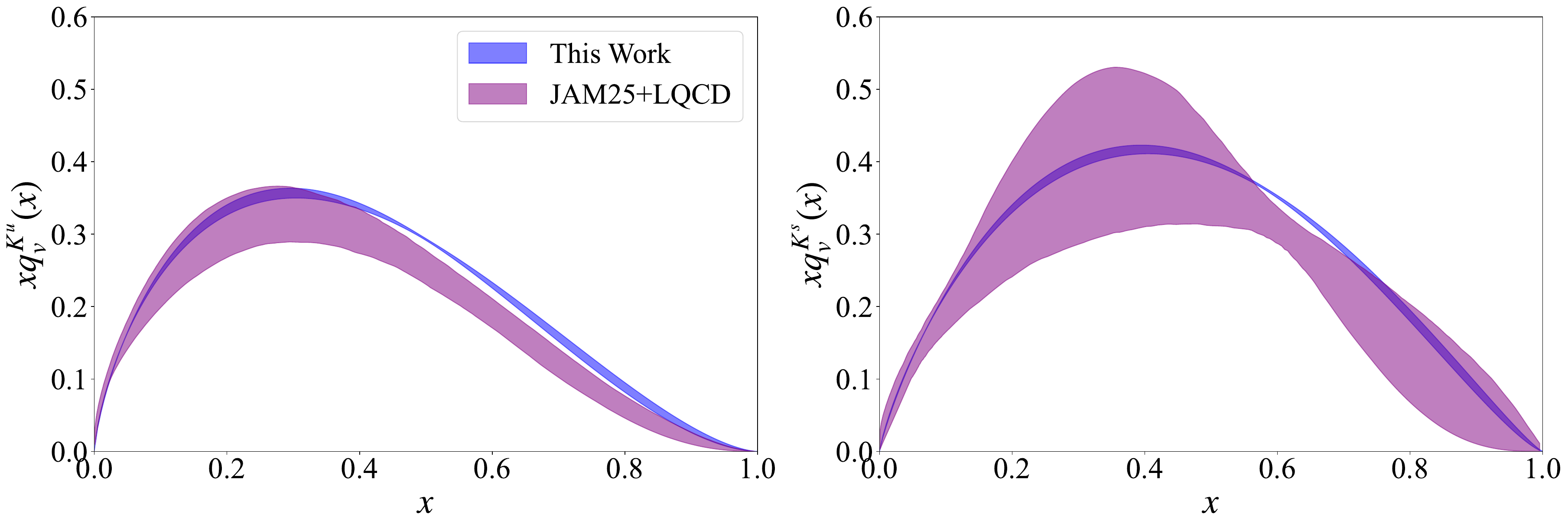}
    \vspace*{-0.35cm}    
    \caption{$xq_v$ for the kaon up- (left) and strange-quark (right) from this work (blue band) compared to the JAM 2025 analysis with lattice-QCD constraints~\cite{Barry:2025wjx} (purple band) at $\mu = 5.2$ GeV.}
    \label{fig:5.2GeV_kaon_comparison}
\end{figure}

\section{SU(3) flavor symmetry breaking}
\label{sec:su3}

An important aspect of comparing the pion and kaon results is the investigation of SU(3) flavor symmetry breaking arising from the presence of the heavier strange quark in the kaon. In the exact SU(3)-flavor symmetric limit, where the quark masses are degenerate, the pion and kaon PDFs are related through flavor symmetry, implying identical Mellin moments after appropriate flavor interchange. Thus, deviations between the corresponding moments provide direct information on the modification of the Mellin moments (and the PDF) induced by the strange-quark mass.

\begin{figure}[!h]
    \centering
    \includegraphics[width=0.9\linewidth]{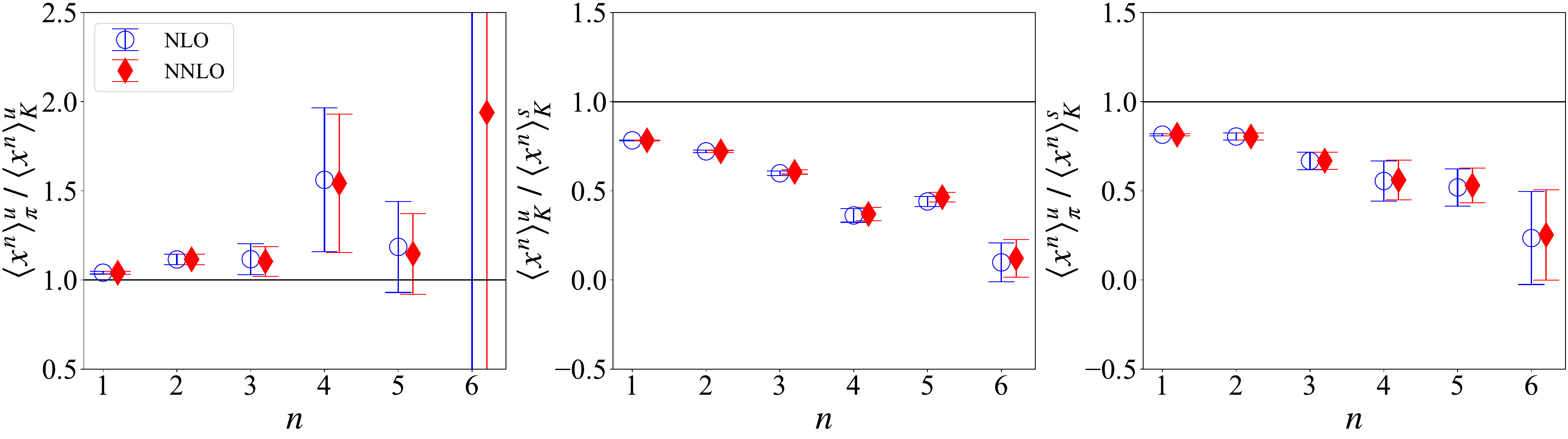}
    \vspace*{-0.45cm}
    \caption{Ratios of same order Mellin moments for pion to kaon up (left), kaon up to kaon strange (middle), and pion to kaon strange (left) at NLO (blue) and NNLO (red) accuracy.}
    \label{fig:SU3}
\end{figure}

To quantify these effects, we take ratios of same-order Mellin moments extracted for the pion and kaon within the same analysis framework. In particular, we compare the ratios $\langle x^n\rangle_\pi^u/\langle x^n\rangle_K^u$, $\langle x^n\rangle_K^u/\langle x^n\rangle_K^s$, and $\langle x^n\rangle_\pi^u/\langle x^n\rangle_K^s$, using both NLO and NNLO Wilson coefficients, as shown in Fig.~\ref{fig:SU3}. The results show that the up-quark moments of the pion and kaon are similar. We note that the errors for $n>3$ increase significantly in the ratio, and the ratio is inconclusive. The ratio $\langle x\rangle_\pi^u/\langle x\rangle_K^u$ differs from unity by only about $4\%$, and by about 10\% for both $\langle x^2\rangle_\pi^u/\langle x^2\rangle_K^u$ and $\langle x^3\rangle_\pi^u/\langle x^3\rangle_K^u$.
A much stronger flavor dependence is observed in the comparison of the kaon strange-quark moments with those of the up quark. For the first moment, we find $\langle x\rangle_K^u/\langle x\rangle_K^s \approx 0.8$, corresponding to an SU(3)-breaking effect at the level of roughly $20$ - $25\%$. The deviation from unity becomes increasingly pronounced for higher moments, reaching approximately $40$ - $60\%$ for $\langle x^3\rangle$ - $\langle x^5\rangle$. A similar trend is observed for the ratio $\langle x^n\rangle_\pi^u/\langle x^n\rangle_K^s$. We emphasize that the interpretation of the ratios becomes less quantitative for the highest moments. Although the central values show an increasing deviation from unity as $n$ increases, the uncertainties of the higher moments are also significantly larger, as discussed in Sec.~\ref{sub:final_results}. Therefore, the most robust conclusion comes from the first few moments. Overall, the observed pattern provides clear evidence of SU(3)-flavor symmetry breaking in meson structure and is qualitatively consistent with previous lattice investigations and phenomenological studies of pion and kaon PDFs (see, e.g., Refs.~\cite{Alexandrou:2020gxs,Alexandrou:2021mmi,Alexandrou:2021ztx}).

\section{Conclusions and Outlook}
\label{sec:conclusions}

In this work, we presented a lattice-QCD determination of the Mellin moments of the unpolarized pion and kaon PDFs using matrix elements of nonlocal quark bilinear operators. The calculation is based on the same matrix elements used in our previous determination of the pion and kaon PDFs~\cite{Miller:2025wgr}. Here we employed the short-distance factorization framework and the operator product expansion to extract the Mellin moments from coordinate-space matrix elements, with the meson momentum boost up to $P_3=2.07$~GeV. The analysis has been carried out using an $N_f=2+1+1$ ensemble with maximally twisted-mass fermions with a pion mass of 260 MeV.

By construction, the methodology we used provides access to several moments within a unified fit framework and avoids the need to construct separate local operators for each moment. We employed and compared two fitting strategies for extracting the moments. The first approach analyzes the reduced Ioffe-time distribution at fixed values of $z$, providing insight into how the extracted moments depend on the individual Wilson-line separation and enabling a direct assessment of residual higher-twist and discretization effects. The second approach performs a combined fit in $(P_3,z)$ space, significantly increasing the number of degrees of freedom and leading to a more stable determination of the moments. We found that the combined-fit strategy provides improved stability and better control over systematic effects, particularly for the higher moments.

We investigated several sources of systematic uncertainty. First, we studied the dependence on the truncation order $n_{\rm max}$ of the OPE and found that $n_{\rm max}=5$ (for odd moments) and $n_{\rm max}=6$ (for even moments) provides a suitable balance between stability and sensitivity to higher moments. Second, we examined the dependence on the fit window in $z$, using several choices of $z_{\rm min}$ and $z_{\rm max}$. The lowest moments are comparatively stable under these variations, whereas the higher moments are more sensitive to the fit range, as expected given their stronger dependence on the large-Ioffe-time behavior of the matrix elements. Third, we compared the results obtained with NLO and NNLO Wilson coefficients. The NNLO corrections lead to small shifts relative to NLO, with the effect most visible for the lowest moments.

We also explored an RG-improved analysis in which the matching is performed at an initial scale and the resulting moments are evolved to $\mu=2$ GeV. This provides an important consistency check on the perturbative treatment of the lowest moments. The analysis that includes evolution shows improved perturbative stability at NNLOevo compared with NLOevo, and the results are broadly consistent with the fixed-scale extraction. However, the short-distance region required for reliable RG evolution provides limited sensitivity to the higher moments. For this reason, we use the RG-improved analysis as a perturbative consistency check rather than the final analysis.

Our final results are taken from the NNLO combined $(P_3,z)$ fits with $n_{\rm max}=6$, using weighted averages over the fit windows with $z_{\rm min}\in[3a,4a]$ and $z_{\rm max}\in[4a,6a]$. The quoted uncertainties separate statistical effects, the systematic dependence on the coordinate-space fit window, and the difference between the NLO and NNLO central values. We find that the first few moments are well-constrained, whereas higher moments become increasingly sensitive to the fit window. This behavior is expected, since higher moments receive enhanced support from the large-$x$ region of the PDF and are therefore more difficult to determine from the finite Ioffe-time range available in the lattice data.
We compare the extracted Mellin moments up to $\langle x^4 \rangle$ with other lattice-QCD studies using non-local operators, and we find similar features for all moments. Deviations among all data are more pronounced in $\langle x^2 \rangle$. 

We also reconstruct the valence PDFs using a normalized two-parameter ansatz constrained by the extracted moments. Comparing fits that include moments up to $n_{\rm max}=4$ and $6$ shows that the broad features of the distributions are similar, although the detailed shape remains sensitive to the highest moments included. Comparisons with phenomenological determinations and experimental data, where available, show an overall agreement for the pion, kaon, as well as for the ratio $q_K^u(x)/q_\pi^u(x)$.

The ratios of same-order moments provide a direct measure of SU(3)-flavor-symmetry breaking. The pion and kaon up-quark moments differ only mildly, by about $4\%$ for $\langle x\rangle$ and approximately $10\%$ for $\langle x^2\rangle$ and $\langle x^3\rangle$. In contrast, the kaon strange-quark moments are substantially larger than the corresponding up-quark moments. For the first moment, the ratio $\langle x\rangle_K^u/\langle x\rangle_K^s$ is approximately $0.8$, corresponding to a breaking effect of about $20$ - $25\%$, with the deviation from unity increasing for higher moments. 

An important outcome of this work is that it further demonstrates the utility of nonlocal matrix elements for extracting several Mellin moments within a single framework, without the need to construct a separate local operator with a higher covariant derivative for each moment. This builds on earlier developments using OPE-based analyses of nonlocal operators, while extending such studies to the pion and kaon with a detailed assessment of the fit-window dependence, perturbative-order effects, and moment reconstruction. In traditional lattice-QCD approaches based on local operators, extracting higher moments becomes increasingly challenging due to operator mixing, power-divergent renormalization, and rapidly deteriorating signal-to-noise ratios. The nonlocal-operator strategy offers a complementary path, using the same matrix elements that underlie quasi- and pseudo-distribution studies. Although the higher moments obtained here remain statistically limited and susceptible to systematic effects, the results provide a useful benchmark for the method and guidance for future calculations.

\begin{acknowledgements}

J. M. and M.~C. acknowledge financial support by the U.S. Department of Energy, Office of Nuclear Physics,  under Grant No.\ DE-SC0025218.
J.D received support from Argonne National Laboratory under the contract ``Pion and Kaon Form Factors using Lattice QCD''.
J. Torsiello acknowledges support by the U.S. Department of Energy, Office of Science, Office of Advanced Scientific Computing Research, Department of Energy Computational Science Graduate Fellowship under Award Number DE-SC0024386~\footnote{This manuscript was prepared as an account of work sponsored by an agency of the United States Government. Neither the United States Government nor any agency thereof, nor any of their employees, makes any warranty, express or implied, or assumes any legal liability or responsibility for the accuracy, completeness, or usefulness of any information, apparatus, product, or process disclosed, or represents that its use would not infringe privately owned rights. Reference herein to any specific commercial product, process, or service by trade name, trademark, manufacturer, or otherwise does not necessarily constitute or imply its endorsement, recommendation, or favoring by the United States Government or any agency thereof. The views and opinions of authors expressed herein do not necessarily state or reflect those of the United States Government or any agency thereof.}
K.~C.\ is supported by the National Science Centre (Poland) grant OPUS No.\ 2021/43/B/ST2/00497. 
The authors acknowledge partial support from the U.S. Department of Energy, Office of Science, Office of Nuclear Physics, under the umbrella of the Quark-Gluon Tomography (QGT) Topical Collaboration, with Award DE-SC0023646.
Computations for this work were carried out in part on facilities of the USQCD Collaboration, which are funded by the Office of Science of the U.S. Department of Energy. 
This research includes calculations carried out on HPC resources supported in part
by the National Science Foundation through major research instrumentation grant number 1625061 and by the US Army Research Laboratory under contract number W911NF-16-2-0189. This research used resources of the Oak Ridge Leadership Computing Facility, which is a DOE Office of Science User Facility supported under Contract DEAC05-00OR22725.
This research used resources of the National Energy Research
Scientific Computing Center, a DOE Office of Science User Facility
using NERSC award ALCC-ERCAP0030652.
This research used resources of the National Energy Research
Scientific Computing Center, a DOE Office of Science User Facility
supported by the Office of Science of the U.S. Department of Energy
under Contract No. DE-AC02-05CH11231 using NERSC award
NP-ERCAP0037566.
The gauge configurations have been generated by the Extended Twisted Mass Collaboration on the KNL (A2) Partition of Marconi at CINECA, through the Prace project Pra13\_3304 ``SIMPHYS".
Inversions were performed using the DD-$\alpha$AMG solver~\cite{Frommer:2013fsa} with twisted mass support~\cite{Alexandrou:2016izb}. 
\end{acknowledgements}

\appendix
\section{NLO Wilson Coefficient Derivation}
\label{appA}

In the main text, the extraction of the Mellin moments relies on the short-distance expansion of the reduced Ioffe-time distribution, Eq.~\eqref{eqn:moment_expansion}. The Wilson coefficients entering that expression are obtained from the perturbative matching between the pseudo-Ioffe-time distribution and the light-cone Ioffe-time distribution. In this appendix, we outline the derivation of the NLO coefficient used in the fixed-order analysis of Sec.~\ref{sub:z_fixed}. This derivation also fixes the normalization and plus-prescription conventions used throughout the manuscript.

At next-to-leading order (NLO), the relation between the pseudo-Ioffe time distribution and the light-cone Ioffe time distribution is given as,
\begin{equation}
    \mathcal{M}(\nu,z^2) = \mathcal{Q}(\nu,\mu^2)+ \frac{\alpha_sC_F}{2\pi}\int_0^1 du~\left[\ln\left(z^2\mu^2\frac{e^{2\gamma_E +1}}{4}\right) B(u) + L(u)\right] \mathcal{Q}(u\nu,\mu^2),
    \label{eqn: Matching_Kernel}
\end{equation}
where $B(u)$ and $L(u)$ are the evolution and matching plus-prescription kernels respectively. They are given by
\begin{equation}
    B(u) = \left[ \frac{1+u^2}{u-1}\right]_+
    \label{eqn: Evolution}
\end{equation}
\begin{equation}
    L(u) = \left[4\frac{\ln(1-u)}{u-1} - 2(u-1) \right]_+\,.
    \label{eqn: Matching}
\end{equation}
In order to relate $\mathcal{M}(\nu,z^2)$ to the Mellin moments, we take note that $\mathcal{Q}(\nu,\mu^2)$ is the Fourier transform of the light-cone PDF. That is,
\begin{equation}
    \mathcal{Q}(\nu,\mu^2) = \int_{-1}^1 dx~e^{i\nu x}q(x) = \sum_{n=0}^{\infty} \frac{(i\nu)^n}{n!}\langle x^n\rangle\,.
    \label{eqn: ITD_FT}
\end{equation}
Substituting Eqs.~\eqref{eqn: Evolution} - \eqref{eqn: ITD_FT} into Eq.~\eqref{eqn: Matching_Kernel}, and also taking note that the plus prescription definition is given as $\int_0^1 du~[h(u)]_+g(u\nu) = \int_0^1du~h(u)[g(u\nu)-g(\nu)]$, we get
\begin{equation}
    \mathcal{M}(\nu,z^2) = \sum_{n=0}^{\infty}\frac{(i\nu)^n}{n!} \langle x^n\rangle \left[1 + 
    \frac{\alpha_sC_f}{2\pi}\int_0^1du~\left[\ln\left(z^2\mu^2\frac{e^{2\gamma_E + 1}}{4}\right)\left(\frac{1+u^2}{u-1}\right) +4\frac{\ln(1-u)}{u-1}-2(u-1)\right](u^n-1)  \right]\,.
\end{equation}
Upon solving these integrals, we end up with a closed form of the pseudo-Ioffe time distribution,
\begin{equation}
    \mathcal{M}(\nu,z^2) = \sum_{n=0}^{\infty}\frac{(i\nu)^n}{n!} \langle x^n\rangle\left[1 + \frac{\alpha_sC_F}{2\pi}\left[\ln\left(z^2\mu^2 \frac{e^{2\gamma_E+1}}{4}\right)\left(H_n+H_{n+2}-\frac{3}{2}\right)  + 2\left(\frac{1}{n+1} - \frac{1}{n+2}-H_n^2 -H_n^{(2)}\right)-1 \right] \right]\,,
    \label{eqn:Our_expansion}
\end{equation}
where $H_n^{(i)}$ are the generalized harmonic numbers, given by
\begin{equation}
    H_n^{(i)} = \sum_{j=1}^n \frac{1}{j^i}\,,
\end{equation}
and $H_n^{(1)} \equiv H_n$. In order to calculate the NNLO Wilson coefficients, a similar process is done numerically for the NNLO matching kernel at fixed values of $n$.

It is important to note that the choice of Wilson Coefficient is similar to literature, but written differently, such as  \cite{Gao:2022iex,Holligan:2024umc}. In these works, the ratio of bare matrix elements is written as an infinite sum in both the numerator and denominator, where we will denote their Wilson coefficients as $C_n$,
\begin{equation}
    \mathcal{M}(\nu,z^2) = \frac{\sum_{n=0} C_n\frac{(izP_z)^n}{n!}\langle x^n\rangle}{\sum_{n=0} C_n\frac{(izP_z^0)^n}{n!}\langle x^n\rangle}~.
    \label{eqn:Other_lit_expansion}
\end{equation}
It is given that $C_n$ is
\begin{equation}
    C_n = 1 + \frac{\alpha_sC_F}{2\pi}\left[\left(\frac{3+2n}{2+3n+n^2}+2H_n \right)\ln\left(\frac{\mu^2z^2e^{2\gamma_E}}{4}\right) + \frac{5+2n}{2+3n+n^2} +2(1-H_n)H_n -2H_n^{(2)} \right]~.
\end{equation}

In the limit of $P_z^0\rightarrow 0$ equation \ref{eqn:Other_lit_expansion}, becomes, along with $\langle x^0\rangle = 1$ in the forward limit,
\begin{equation}
    \mathcal{M}(\nu,z^2) = \sum_{n=0}^{\infty}\frac{C_n}{C_0}\frac{(i\nu)^n}{n!} \langle x^n\rangle ~.
\end{equation}
Focusing only at NLO accuracy, we note that we can calculate $C_0$, giving
\begin{equation}
    \frac{C_n}{C_0} = \frac{1 + \frac{\alpha_sC_F}{2\pi}\left[\left(\frac{3+2n}{2+3n+n^2}+2H_n \right)\ln\left(\frac{\mu^2z^2e^{2\gamma_E}}{4}\right) + \frac{5+2n}{2+3n+n^2} +2(1-H_n)H_n -2H_n^{(2)} \right]}{1 + \frac{\alpha_sC_F}{2\pi}[\frac{3}{2}\ln(\frac{\mu^2z^2e^{2\gamma_E}}{4})+\frac{5}{2}]}~.
\end{equation}
In order to show $C_n/C_0 = c_n$ of Eq.~\eqref{eq:expansion}, we Taylor expand the denominator, collecting up to $\mathcal{O}(\alpha_s)$ terms, leading to 
\begin{equation}
    \frac{C_n}{C_0} = 1 + \frac{\alpha_sC_F}{2\pi}\left[\left(\frac{3+2n}{2{+}3n{+}n^2} {+}2H_n {-} \frac{3}{2} \right)\ln\left(\frac{\mu^2z^2e^{2\gamma_E}}{4}\right) + \frac{5+2n}{2{+}3n{+}n^2}+2(1{-}H_n)H_n-2H_n^{(2)}-\frac{5}2{} \right] = c_n
\end{equation}

\section{Evolution Relations}
\label{appB}

The RG-improved analysis of Sec.~\ref{sub:evol} requires evolving the Wilson coefficients from an initial scale $\mu_0$, associated with the coordinate-space separation $z$, to the reference scale $\mu=2~{\rm GeV}$. In this appendix, we collect the evolution relations used in that analysis. These expressions make explicit how the scale dependence is controlled by the QCD beta function and the nonsinglet anomalous dimensions (see Table~\ref{tab:perturbative_values}), and they provide the perturbative ingredients needed for the NLOevo and NNLOevo comparisons presented in the main text.
To find the LL (NLOevo) and NLL (NNLOevo) relations, we utilize the fact that the Wilson coefficients transform by,
\begin{equation}
    \ln\left(\frac{c_n(\mu)}{c_n(\mu_0)}\right) = \int_{\alpha_s(\mu_0)}^{\alpha_s(\mu)} d(\alpha_s(\mu')) \frac{\gamma_n(\alpha_s(\mu'))}{\beta(\alpha_s(\mu'))}\,.
\end{equation}
This relation is dependent upon $\beta(\alpha_s)$, which is given by
\begin{equation}
    \beta(\alpha_s) = -\alpha_s \sum_{n=0}^{\infty} \left(\frac{\alpha_s}{4\pi}\right)^{n+1}\beta_n\,,
    \label{eqn:beta}
\end{equation}
and the nonsinglet anomalous dimension $\gamma_n$, which is defined to be the negative of the Mellin moment of the non-singlet splitting function \cite{CURCI198027},
\begin{equation}
    \gamma_n(\alpha_s) = -\int_{-1}^1dx~x^nP_{\mathrm{ns}}(x) =\sum_{i=0}^{\infty} \left(\frac{\alpha_s}{4\pi}\right)^{i+1}\gamma_n^{(i)}\,.
    \label{eqn:AD}
\end{equation}

The LL relation can be derived by truncating after the first element in Eq.~\eqref{eqn:beta} and Eq.~\eqref{eqn:AD}. The NLL relation is truncated after the second term. The LL relation is found to be
\begin{equation}
    \ln\left(\frac{c_n(\mu)}{c_n(\mu_0)} \right) = \int_{\alpha_s(\mu_0)}^{\alpha_s(\mu)} d(\alpha_s(\mu')) \frac{\frac{\alpha_s}{4\pi}\gamma_n^{(0)}}{-\frac{\alpha_s^2}{4\pi}\beta_0} = -\frac{\gamma_n^{(0)}}{\beta_0}\ln\left(\frac{\alpha_s(\mu)}{\alpha_s(\mu_0)}\right)\,,
\end{equation}
and the NLL relation is found to be
\begin{equation}
    \ln\left(\frac{c_n(\mu)}{c_n(\mu_0)} \right) = \int_{\alpha_s(\mu_0)}^{\alpha_s(\mu)} d(\alpha_s(\mu')) \frac{\frac{\alpha_s}{4\pi}\gamma_n^{(0)} + \frac{\alpha_s^2}{(4\pi)^2}\gamma_n^{(1)}}{-\frac{\alpha_s^2}{4\pi}\beta_0 -\frac{\alpha_s^3}{(4\pi)^2}\beta_1} = -\frac{\gamma_n^{(0)}}{\beta_0}\ln\left(\frac{\alpha_s(\mu)}{\alpha_s(\mu_0)}\right) - \left(\frac{\gamma_n^{(1)}}{\beta_1} - \frac{\gamma_n^{(0)}}{\beta_0}\right)\ln\left(\frac{4\pi\beta_0+\beta_1\alpha_s(\mu)}{4\pi\beta_0+\beta_1\alpha_s(\mu_0)} \right)\,.
\end{equation}

Similar to Appendix \ref{appA}, it is important to note that the choices of the final result of what is called the anomalous dimension $\gamma_n$ in this work, is related to that of other literature such as Ref. \cite{Gao:2022iex} (which we denote here as $\Gamma_n$). Focusing only on the LL (NLOevo) coefficients, 
\begin{equation}
    \frac{C_n(\mu)}{C_0(\mu)} = \frac{C_n(\mu_0)}{C_0(\mu_0)} \frac{\left(\frac{\alpha_s(\mu)}{\alpha_s(\mu_0)}\right)^{-\Gamma_n^{(0)}/\beta_0}}{\left(\frac{\alpha_s(\mu)}{\alpha_s(\mu_0)}\right)^{-\Gamma_0^{(0)}/\beta_0}} = \frac{C_n(\mu_0)}{C_0(\mu_0)} \left(\frac{\alpha_s(\mu)}{\alpha_s(\mu_0)}\right)^{\frac{-\Gamma_n^{(0)}+\Gamma_0^{(0)}}{\beta_0}}~.
    \label{eqn:Other_lit_evolution}
\end{equation}

Having already shown the equivalence of $\frac{C_n(\mu_0)}{C_0(\mu_0)} = c_n(\mu_0)$ in Appendix \ref{appA}, we shift our focus on the anomalous dimension. In Ref. \cite{Gao:2022iex}, it is defined as
\begin{equation}
    \Gamma_n^{(0)} = \frac{3C_F}{2}-\int_0^1 dx ~x^nP_{qq}^{(0)}(x)~,
\end{equation}
 where at $n=0$, $\Gamma_0^{(0)} = \frac{3C_F}{2}$ and at 1-loop $P_{\mathrm{ns}}(x)$ becomes the QCD s. From this, the exponent of Eq.~\eqref{eqn:Other_lit_evolution} becomes
\begin{equation}
    -\Gamma_n^{(0)}+\Gamma_0^{(0)} = \int_0^1 dx~x^nP_{qq}^{(0)}(x) = -\gamma_n^{(0)}~,
\end{equation}
which leads to the result that $\frac{C_n(\mu)}{C_0(\mu)} = c_n(\mu)$.

\begin{table}[h!]
\begin{center}
\renewcommand{\arraystretch}{1.2}
\begin{tabular}{l|ccccccc}
\hline
$n$ & 0 & 1 & 2 & 3 & 4 & 5 & 6  \\ \hline
$\beta_n$ & $8.33333$ & $51.3333$ & $406.352$ & - & - & - & - \\\hline
$\gamma_n^{(0)}$ & 0 & 3.55556 & 5.55556 & 6.97778 & 8.08889 & 9.00317 & 9.78095 \\\hline
$\gamma_n^{(1)}$ & 0 & 34.4085 & 50.3909 & 58.7936 & 67.4518 & 72.2227 & 78.6661 \\\hline
\end{tabular}
\caption{Values for $\beta_n$, $\gamma_n^{(0)}$, and $\gamma_n^{(1)}$ used in this work to calculate $\alpha_s$ and DGLAP evolution when $n\in[1,6]$ and  $N_f=4$.}
\label{tab:perturbative_values}
\end{center}
\end{table}

\bibliography{references.bib}

\end{document}